\documentclass[aps,twocolumn,prl,showpacs,superscriptaddress,groupedaddress]{revtex4}  
\usepackage{dcolumn}
\usepackage{bm}
\usepackage{color}

\def\ltape{\hbox{\ $<$\hskip -8pt\raise -4pt\hbox{$\sim$}\ }}
\def\gtape{\hbox{\ $>$\hskip -8pt\raise -4pt\hbox{$\sim$}\ }}

   \ifx\pdfoutput\undefined
   \usepackage[dvips]{graphicx}
   \else
   \usepackage[pdftex]{graphicx}
   \fi
   \usepackage{epstopdf}
   \DeclareGraphicsExtensions{.pdf,.gif,.jpg,.eps,.png}
  
   \usepackage{amsmath}
  \usepackage{amssymb}

\pdfoutput=1

\begin{document}

\title{Orientation and stability of asymmetric magnetic reconnection x-line}

\author{Yi-Hsin~Liu}
\affiliation{Dartmouth College, Hanover, NH 03750}
\author{Michael~Hesse}
\affiliation{University of Bergen, Bergen, Norway}
\affiliation{Southwest Research Institute, San Antonio, TX 78238}
\author{Tak Chu~Li}
\affiliation{Dartmouth College, Hanover, NH 03750}
\author{Masha~Kuznetsova}
\affiliation{NASA-Goddard Space Flight Center, Greenbelt, MD 20771}
\author{Ari Le}
\affiliation{Los Alamos National Laboratory, Los Alamos, NM 87545}
\date{\today}
\begin{abstract}
The orientation and stability of the reconnection x-line in asymmetric geometry is studied using three-dimensional (3D) particle-in-cell simulations. We initiate reconnection at the center of a large simulation domain to minimize the boundary effect. The resulting x-line has sufficient freedom to develop along an optimal orientation, and it remains laminar. Companion 2D simulations indicate that this x-line orientation maximizes the reconnection rate. The divergence of the non-gyrotropic pressure tensor breaks the frozen-in condition, consistent with its 2D counterpart. We then design 3D simulations with one dimension being short to fix the x-line orientation, but long enough to allow the growth of the fastest growing oblique tearing modes. This numerical experiment suggests that reconnection tends to radiate secondary oblique tearing modes if it is externally (globally) forced to proceed along an orientation not favored by the local physics. The development of oblique structure easily leads to turbulence inside small periodic systems.
\end{abstract}


\maketitle

\subsection{1. Introduction}
Magnetic reconnection plays the critical role in the plasma transport and magnetic energy release at Earth's magnetopause, the sharp boundary separating Earth's magnetosphere and solar wind plasmas. To understand the global convection of plasmas and magnetic flux around Earth, it is imperative to know {\it where} reconnection will take place on this boundary layer \citep{dungey61a}. With purely southward interplanetary magnetic fields (IMF) in the solar wind, it is clear that the dayside reconnection will occur along the equatorial plane. The resulting locus that connects these reconnection locations is called the {\it reconnection-line} [e.g.,\citep{trattner04a}]. However, the location and orientation of the reconnection-line become less clear when the IMF points in a clock-angle different than southward (i.e., the Sun-Earth direction is the rotation axis). Observations suggested a``tilted'' reconnection-line in this situation \citep{dunlop11a,phan06a,pu07a,scurry94a,trattner07a,fear12a,dunlop11b,wild07a,kawano05a,daly84a}. A similarly tilted reconnection-line was illustrated in global MHD simulations by tracing the global magnetic separator \citep{komar13a}. Predictions of the reconnection-line location on the magnetopause were previously made by mapping out locations that maximize local quantities, such as the shear angle \citep{trattner07a}, current density \citep{alexeev98a}, and Poynting flux divergence \citep{papadopoulos99a}. Another approach was based on the vacuum superposition of Earth dipolar and solar wind magnetic fields \citep{cowley73a,siscoe01a,dorelli07a}. 

\begin{figure}
\includegraphics[width=8cm]{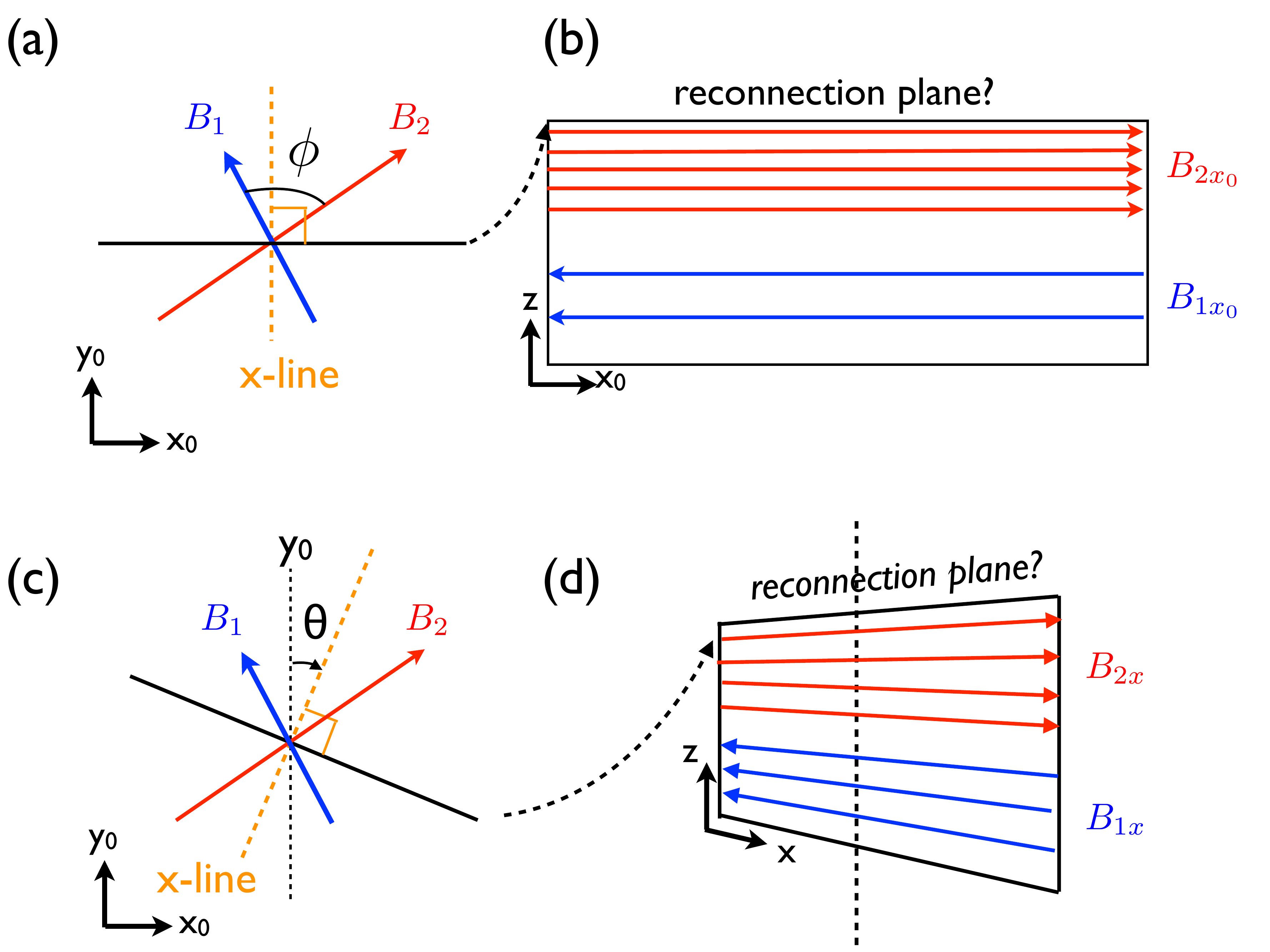} 
\caption {Illustration of the question to solve.
}
\label{question}
\end{figure}

In this work, we approach this problem from the local aspect of reconnection by studying the orientation of the reconnection x-line in slab geometry. Even with this simplified geometry, understanding this 3D nature of magnetic reconnection is already challenging and a strict theoretical treatment does not exist. Researchers have used the same principle that determines the local x-line orientation to map out the reconnection location on the global magnetopause \citep{komar15a}.  The result of this study suggests that the tangent of a global reconnection-line will eventually align with the local x-line orientation.
The question to solve and the coordinate system employed in this study are further illustrated in Fig.~\ref{question}; Magnetic fields on two sides of the boundary layer (like Earth's magnetopause) can shear at an arbitrary angle $\phi$. Here we consider the boundary normal to the $z$-direction and the $B_z$ component to be negligible. If we take a 2D cut depicted by the black line in Fig.~\ref{question}(a), the in-plane component of magnetic fields on the two sides are anti-parallel as illustrated in (b), and thus reconnection can occur on this plane. However, this 2D plane is not the only possible choice. We take another 2D plane in (c), and there are also in-plane anti-parallel magnetic fields for reconnection as illustrated in (d) although the in-plane field strength changes on this plane. Therefore the question to ask is, given different magnetic field and plasma conditions on two sides of the current sheet, on which plane will reconnection proceed? Since the reconnection x-line (marked by the orange dashed line in (a) and (c)) is always perpendicular to the corresponding 2D reconnection plane, the goal is equivalent to determining the orientation of the x-line. We will quantify the x-line orientation by the angle $\theta$ respected to the ${\bf y}_0$-axis 
(For simplicity, we choose ${\bf y}_0$ to be the direction where the guide field $B_{y0}$ is uniform). 
Hypotheses to this well-defined question were proposed. They include minimizing the in-plane current \citep{sonnerup74a,gonzalez74a}, maximizing the reconnection outflow speed \citep{swisdak07a}, maximizing the reconnection rate \citep{schreier10a,hesse13a, yhliu15b, aunai16a} or maximizing the oblique tearing growth rate \citep{yhliu15b}. 
On the other hand, other than a few studies in literature \citep{schreier10a, yhliu15b} there are not many attempts to study this fundamental nature of magnetic reconnection using first-principle 3D simulations. To resolve the reconnection x-line in the electron-scale will require a fully kinetic description. Thus we use both 3D and 2D particle-in-cell (PIC) simulations to explore this issue. After knowing the optimal orientation favored by the local physics, we further study the response of the system when the x-line is 
forced to misalign with the optimal orientation. This result reveals the potential format of the interplay between the global and local controls.

The structure of this paper is outlined in the following. Section 2 describes the simulation setup. Sec.~3 measures the x-line orientation in the large 3D simulation. Sec.~4 identifies the non-ideal term in Ohm's law that breaks the frozen-in condition. Sec.~5 shows the comparison with companion 2D simulations and theories. Sec.~6 studies the response of the x-line when it is forced to proceed at an orientation not favored by the local physics. Sec.~7 summarizes and discusses our results.\\

\subsection{2. Simulation Setup}
In this paper, kinetic simulations were performed using the electromagnetic particle-in-cell code {\it VPIC} \citep{bowers09a}.  The employed asymmetric current sheet \citep{yhliu15b,hesse13a,aunai13b,pritchett08a} has the magnetic profile, ${\bf B}_0=B_0(0.5+S)\hat{\bf x}_0+B_0\hat{\bf y}_0$ with $S=\mbox{tanh}[(z-3d_i)/L]$, which corresponds to a shear angle $\phi\simeq 82.87^\circ$ across the sheet. This profile gives $B_{2x0}=1.5B_0$ and $B_{1x0}=0.5B_0$ where the subscripts ``1'' and ``2'' correspond to the magnetosheath and magnetosphere sides respectively. The initial current sheet has a half-thickness $L= 0.8 d_i$, and it is shifted from $z=0$ to $z=3d_i$ to accommodate the larger structure expected in the weaker field side;  
the opening angle of the reconnection exhaust boundary on this side should be larger \citep{yhliu18a}. 
The plasma has a density profile $n=n_0[1-(S+S^2)/3]$ that gives $n_2=n_0/3$ and $n_1=n_0$. The uniform total temperature is $T=3B_0^2/(8\pi n_0)$ that consists of contributions from ions and electrons with ratio $T_i/T_e=5$. The mass ratio is $m_i/m_e=25$. The ratio of the electron plasma to gyro-frequency is $\omega_{pe}/\Omega_{ce}=4$ where $\omega_{pe}\equiv(4\pi n_0 e^2/m_e)^{1/2}$ and $\Omega_{ce}\equiv eB_0/m_e c$. In the presentation, densities, time, velocities, spatial scales, magnetic fields and electric fields are normalized to $n_0$, the ion gyro-frequency $\Omega_{ci}$, the Alfv\'enic speed $V_A\equiv B_0/(4\pi n_0 m_i)^{1/2}$, the ion inertia length $d_i\equiv c/\omega_{pi}$, $B_0$ and $V_A B_0/c$, respectively. 

The x-line orientation will be quantified by the angle $\theta$ respect to the ${\bf y}_0$-axis illustrated in Fig.~\ref{question}. A clockwise rotation gives a negative $\theta$.
We can rotate the simulation box along the {\it z}-axis by $\theta_{box}$ so that $\hat{\bf x}=\mbox{cos}\theta_{box}\hat{\bf x}_0-\mbox{sin}\theta_{box}\hat{\bf y}_0$ and $\hat{\bf y}=\mbox{sin}\theta_{box}\hat{\bf x}_0+\mbox{cos}\theta_{box}\hat{\bf y}_0$.
The resulting magnetic field in the new coordinate will be
\begin{equation}
\begin{split}
B_x(z)=B_{x0}(z)\mbox{cos}\theta_{box}+B_{y0}\mbox{sin}\theta_{box}, \\
B_y(z)=-B_{x0}(z)\mbox{sin}\theta_{box}+B_{y0}\mbox{cos}\theta_{box}.
\end{split}
\label{oblique}
\end{equation}
In a 2D system, the orientation of the x-line is fixed in the out-of-plane direction. This machinery allows us to study reconnection at a given x-line orientation $\theta=\theta_{box}$. 

The primary 3D case has $\theta_{box}=0^\circ$, and it has a domain size $L_x \times L_y \times L_z=256d_i \times 256d_i \times 24d_i$ and $4096 \times 4096 \times 384$ cells. The boundary conditions are periodic both in the $x$- and $y$-directions, while in the $z$-direction they are conducting for fields and reflecting for particles. We use $200$ particles per cell. Adopting the methodology in Ref. \cite{yhliu15b}, we localize the perturbation in both the x- and y-directions to initiate reconnection. Companion 2D and 3D simulations with a much shorter {\it y}-extent ($L_y=32d_i$) at a few representative oblique angles $\theta_{box}$ are designed to compare and contrast with the primary 3D case.

\begin{figure}
\includegraphics[width=8cm]{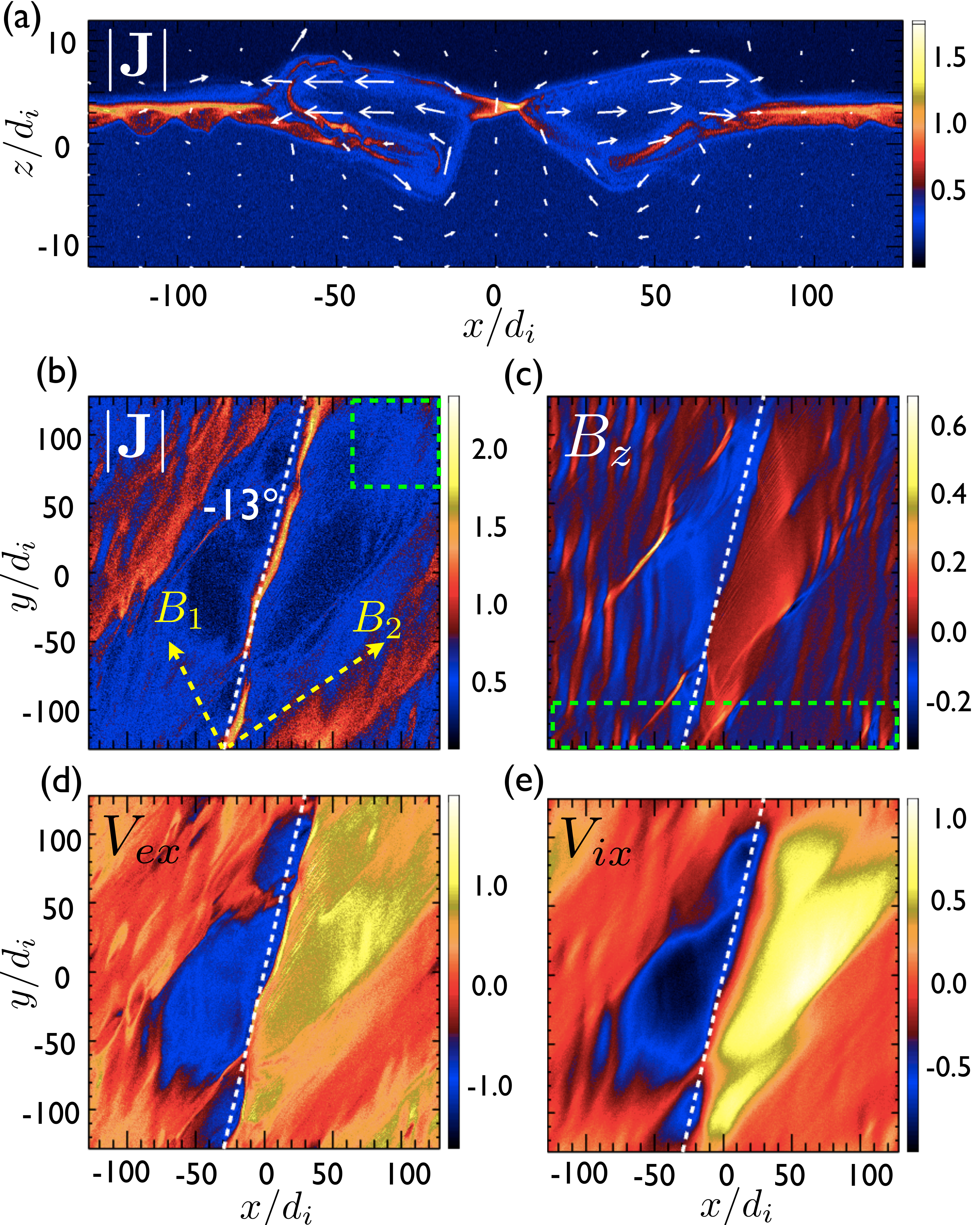} 
\caption {Quantities at time $184/\Omega_{ci}$. In (a) the total current density $|{\bf J}|$ on a 2D plane where $y=0$. The white arrows show the in-plane electron velocities. In (b) the $x$-$y$ cut of $|{\bf J}|$ across the location of the intense current near the x-line. Similarly, in (c) the reconnected field $B_z$, in (d) the electron outflow $V_{ex}$, in (e) the ion outflow $V_{ix}$. On top of the figures, 
yellow arrowed lines in (b) illustrate the magnetic fields on two sides of the current sheet, and white dashed lines in (b)-(e) have $\theta=-13^\circ$. 
}
\label{orientation_3D}
\end{figure}

\subsection{3. X-line Orientation}
Magnetic reconnection is initiated at the center of the simulation domain. 
The reconnection x-line forms and spreads. 
In a slab geometry, a reconnection x-line is best defined by the line of vanishing $B_z$, which is sandwiched between newly generated reconnected field $B_z$. The peak current density also serves as a good proxy to study the x-line orientation when the x-line is quasi-2D \citep{yhliu15b}. 
The total current density $|\bf{J}|$ at $y=0$ and time $184/\Omega_{ci}$ is shown in Fig.~\ref{orientation_3D}(a). To study the orientation of this x-line, we then take the $x-y$ cut of a few quantities across the location of the intense current at $z/d_i \simeq 3.5$. The current density in Fig.~\ref{orientation_3D}(b) captures the distinct x-line that is microscopically narrow but macroscopically long on the $x$-$y$ plane. 
A movie that shows the evolution of $|\bf{J}|$ can be found in the supporting information. 
The x-line in this case is surprisingly laminar and quasi-2D, unlike most 3D simulations where turbulence impacts the current sheet.
The large guide field has suppressed the drift-kink instability \citep{karimabadi03a}. The mild variability of the x-line occurs when the intense current spreads and merges with the current intensified by the background tearing modes at two ends of this primary x-line. 
For reference, the orientations of the asymptotic magnetic fields on both sides are marked by the yellow-dashed arrows. The field strength is proportional to the arrow length. A straight line at orientation $-13^\circ$ is also plotted for comparison. This is the x-line orientation previously determined by the simulation in a $4 \times 4 \times 1.5$ smaller spatial domain (illustrated by the green dashed box at the upper right corner of (b)) and $3$ times shorter evolution time ($60/\Omega_{ci}$) \citep{yhliu15b}. In conjunction with Ref. \cite{yhliu15b}, the comparison demonstrates that this well-defined x-line sustains the same orientation for at least $(184-60)/\Omega_{ci}=124/\Omega_{ci}$, and we do not expect this orientation to change in a larger simulation. While the x-line extent in \cite{yhliu15b} is $\simeq 20 d_i$, the x-line in this larger simulation spread to a spatial length $\simeq 200 d_i$, suggesting that the x-line extent in this regime is purely limited by the system size and there is no intrinsic length limitation in the 3D system. In a slab geometry, the reconnected magnetic field $B_z$ normal to the current sheet most faithfully captures the x-line because it marks the change of the field-line connectivity. The $B_z$ reversal in Fig.~\ref{orientation_3D} shows a similar orientation. The Alfv\'enic flow reversals serve as the strong indicative evidence of magnetic reconnection in in-situ observations [e.g., \citep{burch16a}]. The locus of outflow reversal locations, as captured by $V_{ex}$ and $V_{ix}$ in Fig.~\ref{orientation_3D}(d) and (e), also suggests a similar orientation. Also, note that in (c) the clear stripe structure of $B_z$ arises from the dominant oblique tearing modes that spontaneously grow from the ambient current sheet. The associated plasmoids are observed in (a) for $|x|\gtrsim 75d_i$, outside of the outflow region of the primary reconnection x-line. These stripes make a similar orientation at $-13^\circ$, and this fact has an implication for the x-line stability, that will be discussed later.

\begin{figure}
\includegraphics[width=8cm]{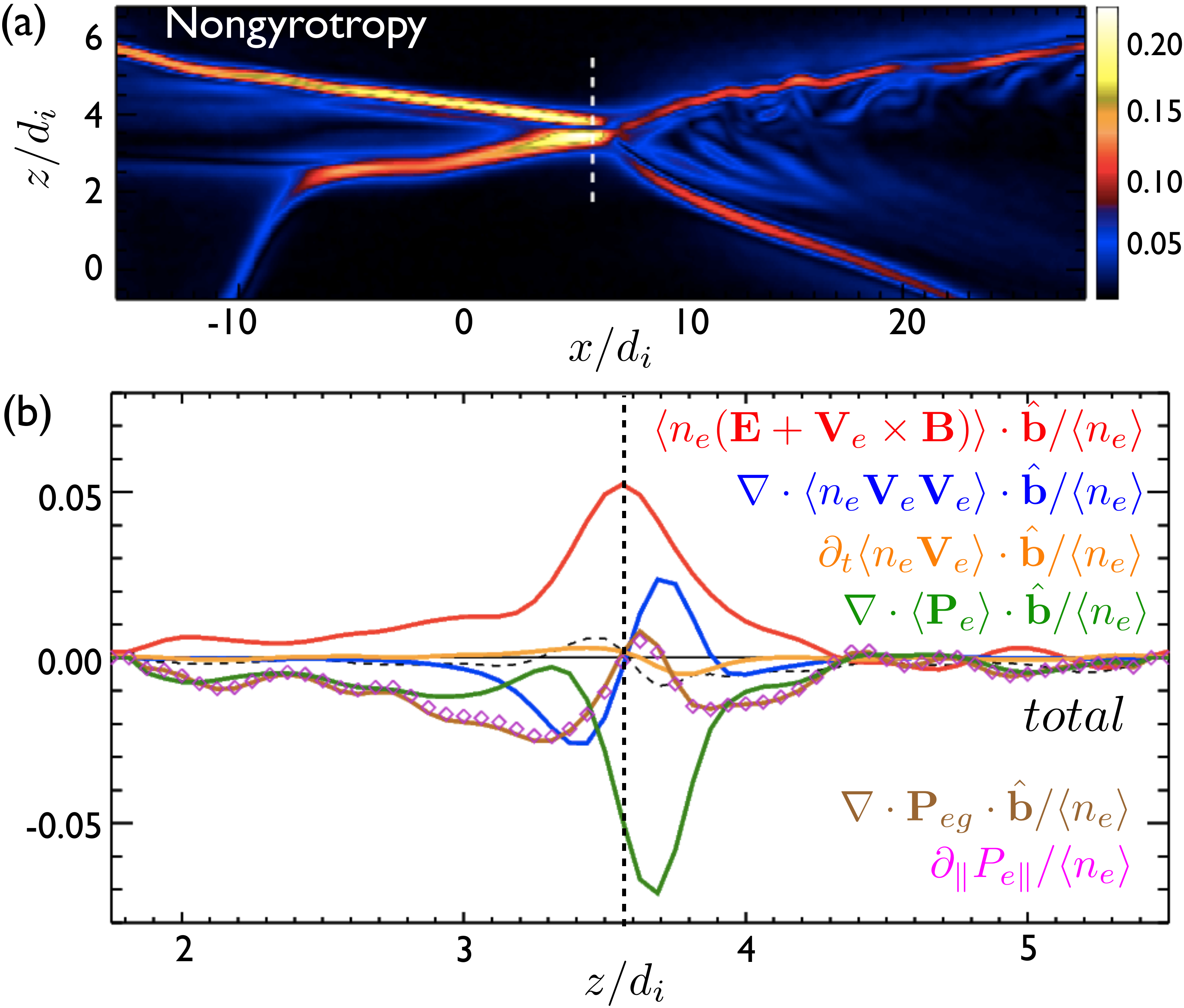} 
\caption { In (a) the measure of the non-gyrotropy ($D_{ng}$) of the pressure tensor. In (b) the decomposition of the non-ideal electric field. The charge $e$ and electron mass $m_e$ are normalized to unity in our presentation.
}
\label{Ohms}
\end{figure}

\subsection{4. Break of the frozen-in condition}
The sharp spatial gradient adjacent to the electron-scale diffusion region makes the ambient plasmas non-gyrotropic. Fig.~\ref{Ohms} (a) shows the non-gyrotropy calculation \citep{aunai13d} $D_{ng}\equiv 2\sqrt{\sum_{i,j} N^2_{ij}}/\mbox{Tr}({\bf P})$ where the non-gyrotropy tensor ${\bf N}={\bf P}-{\bf P}_{eg}$ measures the difference between the full pressure tensor and its gyrotropic approximation. Here ${\bf P}_{eg} \equiv P_{e\perp} {\bf I}+(P_{e\|}-P_{e\perp}) \hat{\bf b} \hat{\bf b}$ with $P_{e\|} \equiv \hat{\bf b}\cdot {\bf P}_e \cdot \hat{\bf b}$ being the electron pressure parallel to the local magnetic field and $P_{e\perp} \equiv [{\rm Tr}({\bf P}_e)-P_{e\|}]/2$ being the pressure perpendicular to the local magnetic field. The intense $D_{ng}$ traces the diffusion region and the sharp outflow exhaust boundaries.
To assess the break of the electron frozen-in condition, we analyze the composition of the non-ideal electric field (along the vertical white dashed line) using the electron momentum equation (i.e., Ohm's law)
\begin{equation}
en_e({\bf E}+{\bf V}_e\times{\bf B}/c)+\nabla\cdot {\bf P}_e + m_e\nabla\cdot \left({n_e {\bf V}_e{\bf V}_e}\right) 
+m_e\frac{\partial}{\partial t}\left(n_e {\bf V}_e\right)=0.
\label{momentum}
\end{equation}
In order to beat the PIC noise in this calculation, it is customary to ensemble average quantities. Since the meaning of ``anomalous dissipations'' arising from an ensemble average (either in a given space extent \citep{che11a,price16a,le17a} or time duration \citep{le17a}) remains unclear \citep{le18a}, here we average the entire equation without further splitting the nonlinear terms into a product of averaged quantities. The ensemble-averaged quantities are marked by the angle bracket in Fig.~\ref{Ohms}(b). $\langle {\bf Q} \rangle\cdot \hat{\bf b}$ indicates that the entire quantity {\bf Q} is time-averaged using 1000 frames within duration $1.7/\Omega_{ci}$, then it is dotted with the averaged unit magnetic vector $\hat{\bf b}\equiv \langle {\bf B}\rangle/\langle B \rangle$. This shows the (time-averaged) quantity in the (time-averaged) parallel direction. 
 
The peak non-ideal electric field ${\bf E}+ {\bf V}_e\times {\bf B}/c$ (red) in the parallel direction is primarily supported by the pressure tensor $\nabla\cdot {\bf P}_e$ (green), and it closely resembles that in the corresponding 2D simulation \citep{hesse16a, QLu13a}. 
To further identify the key contribution in the full pressure tensors, it is useful to evaluate the divergence of its gyrotropic approximation. A similar decomposition is also analyzed in the observations of Magnetospheric Multiscale Mission (MMS) \citep{rager18a,genestreti18a}. 
Outside of the diffusion region, $\nabla\cdot {\bf P}_{eg} \cdot \hat{\bf b}$ (brown) is a good approximation of $\nabla \cdot \langle {\bf P_e}\rangle \cdot \hat{\bf b}$. The contribution from the gyrotropic approximation vanishes near the location of the peak non-ideal electric field, indicating that the primary contribution to the pressure gradient comes from the non-gyrotropy. This is consistent with the idea made by the $D_{ng}$ measurement in Fig.~\ref{Ohms}(a).  
We can further decompose the gyrotropic pressure gradient into $\nabla\cdot {\bf P}_{eg}\cdot \hat{\bf b} =\partial_\| P_{e\|}-(P_{e\|}-P_{e\perp})\partial_\| \ln \langle B\rangle$ where $\partial_\| \equiv \hat{\bf b}\cdot\nabla$. The simulation result suggests that the gyrotropic contribution can be approximated by $\nabla\cdot {\bf P}_{eg}\cdot \hat{\bf b}\approx\partial_\| P_{e\|}$. i.e., the parallel gradient of the parallel component of the pressure tensor (magenta diamonds). The validity of this approximation is also observed in previous 3D simulations \citep{yhliu13a}.

\subsection{5. Companion 2D simulations and theories}
\begin{figure}
\includegraphics[width=8cm]{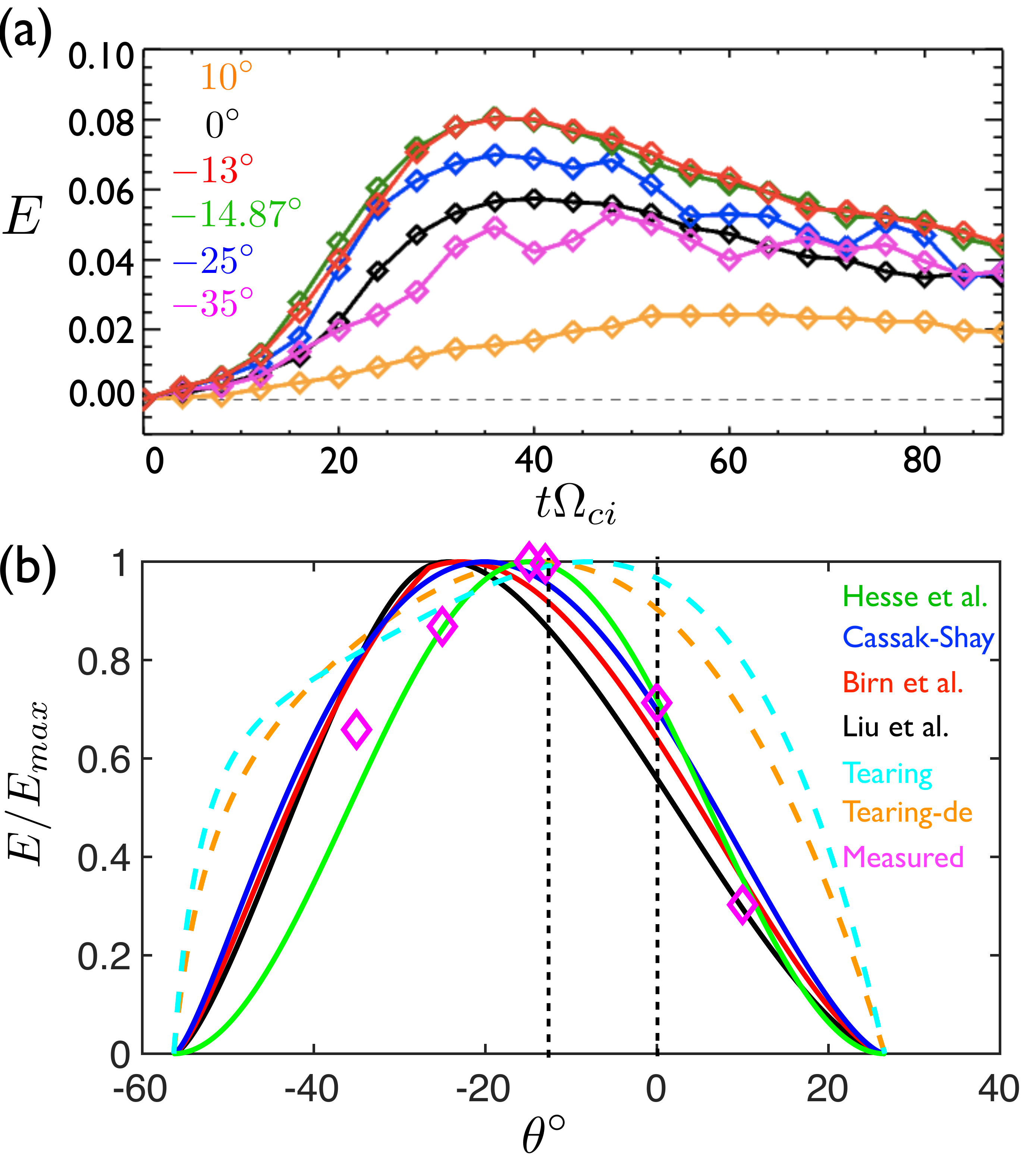} 
\caption {In (a) the evolution of the reconnection rate ($E$) measured on a sample of oblique planes at different $\theta_{box}$. In (b), ``Hesse et al.'', ``Cassak-Shay'', ``Birn et al.'', and ``Liu et al.'' are the predicted reconnection rates from different models. ``Tearing'' shows the tearing growth rate derived in the appendix and ``Tearing-de'' is the modified growth rate in a $d_e$-scale sheet. The measured rates from (a) are plotted as magenta diamonds. Each curve is normalized to its maximum value.
}
\label{orientation_2D}
\end{figure}

Unlike the 3D system where the x-line has sufficient freedom to choose an optimal orientation, in 2D systems the orientation of the x-line is always fixed to the out-of-plane direction because of the translational invariance along this direction. Taking advantage of this artifact, we can study the property of reconnection in a specified orientation. On different oblique planes, the strength of the in-plane magnetic field varies according to Eq.~(\ref{oblique}). 
The in-plane component of magnetic field reverses sign for $\theta_{box}\in [-56.3^\circ, 26.6^\circ]$, and reconnection could operate on any of these oblique planes. In Fig.~\ref{orientation_2D}(a), we show the evolution of reconnection rates on few oblique planes ranging from $\theta_{box}=-25^\circ$ to $10^\circ$. These rates are measured by calculating the change of the in-plane magnetic flux in between the X- and O-points. The measurement suggests that the reconnection rate is maximized at the orientation around $-13^\circ$ (red curve in Fig.~\ref{orientation_2D}(a)), consistent with the orientation manifested in the 3D simulation. This comparison between 3D and 2D systems demonstrates that reconnection proceeds near the maximal reconnection rate. (As an aside, this tendency of maximizing the rate revealed in 3D simulations echoes the hypothesis used to derive the normalized asymmetric reconnection rate 0.1 in recent work \citep{yhliu18a}).

Prompted by this agreement, we now compare our results to the prediction from different rate models. Cassak and Shay \cite{cassak07b} derived an expression of reconnection rate based on conservation laws, $E_{rec}\propto (B_{x1}B_{x2})^{1/2}(B_{x1}+B_{x2})^{-1/2}(B_{x1}\rho_2+B_{x2}\rho_1)^{-1/2}$. Later, Birn et al. \cite{birn10a} included the effect of compression and enthalpy in the calculation. Hesse et al. \cite{hesse13a} proposed that the reconnection rate is proportional to the available magnetic energy based on the reconnecting component $E_{rec}\propto B^2_{x1}B^2_{x2}$, which always leads to a maximal rate at the bisection angle. Recently,  Liu et al. \cite{yhliu18a} modeled the reconnection rate as a function of the opening angle made by the upstream magnetic field. A prediction is attained by maximizing the model rate under the geometrical constraint imposed at the MHD-scale. 
Finally, since the stripe made by the dominant oblique tearing modes (presumably the fastest growing tearing modes) appears to be parallel to the x-line orientation (Fig.~\ref{orientation_3D}(c)), we also derive the growth rate of collisionless oblique tearing modes in the appendix. 
It is not too surprising to see the dominant tearing mode sharing an orientation similar to that of the x-line at its nonlinear state, because a tearing mode is the linear stage of spontaneous reconnection. 
As demonstrated in the next section, the fastest growing oblique tearing becomes active when the x-line is forced to be oriented at an angle different from the optimal orientation.

These predicted reconnection rates are plotted in Fig.~\ref{orientation_2D}(b) as a function of the x-line orientation $\theta$. To facilitate the identification of the optimal angle, each curve is normalized to its maximum. For reference, $\theta =-13^\circ$ (the x-line orientation) and $0^\circ$ (the $y_0$-axis) are marked by the vertical dashed lines. We also plot the peak reconnection rates measured in Fig.~\ref{orientation_2D}(a) as magenta diamonds. The linear growth rates of oblique tearing modes are plotted as a cyan dashed curve. The growth rate based on a thick current sheet maximizes at $\theta \simeq -8^\circ$. However, secondary tearing modes often grow from the non-linear current sheet of $d_e$-scale thickness and the tearing-mode simulation in a $d_e$-scale sheet \citep{yhliu15b} showed the dominant mode with an orientation close to $\theta \simeq-13^\circ$. After accounting for a narrow sheet at $d_e$-scale, a modified theory (also derived in the Appendix-1) is plotted as the orange dashed curve. Two of the closest predictions of the x-line orientation for this case are provided by Hesse et al. \cite{hesse13a} at the bisection angle $\theta \simeq-14.87^\circ$ and the maximum of the modified tearing growth rate at $\theta \simeq-13.8^\circ$. To distinguish which model works better in general will require a thorough parametric study. Nevertheless, these predictions range from $\theta\simeq-8^\circ$ to $-25^\circ$ and are clearly off the $y_0$-axis at $\theta=0^\circ$. The observed x-line orientation falls within this predicted range.

\begin{figure}
\includegraphics[width=9cm]{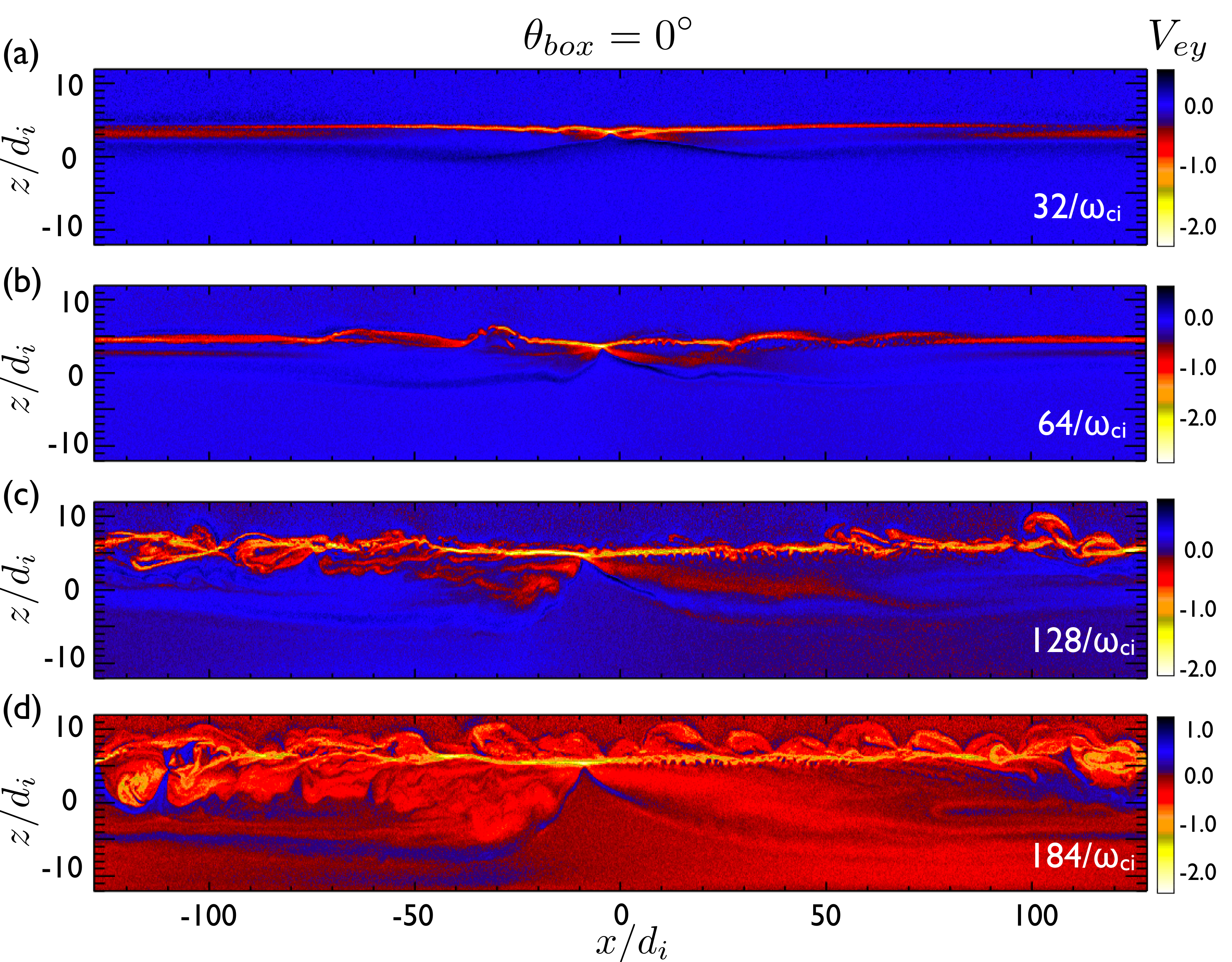} 
\caption {The evolution of reconnection in a companion 3D simulation using $L_y=32d_i$ and $\theta_{box}=0^\circ$. The color shows the electron flow speed in the y-direction ($V_{ey}$).
}
\label{turbulence_evolve}
\end{figure}

\subsection{6. Orientation versus Stability -- a numerical experiment}
At Earth's magnetopause, the initial reconnection-line could be pre-conditioned by the global geometry and external forcing when a relatively planar solar wind touches the bell-shaped magnetosphere at the dayside. The local tangent of such a reconnection-line may not necessarily align with the optimal orientation favored by the local physics. It is thus interesting to explore the stability of reconnection in a 3D system when the x-line does not point to the optimal orientation. As mentioned earlier, when the $L_y$ boundary is extremely short, the quasi-2D system fixes the x-line to the y-direction and completely suppresses any mode that has a finite $k_y$. In the following numerical experiments, we make $L_y$ short to fix the x-line in the $y$-direction but long enough to allow the development of oblique tearing modes, which can spontaneously lead to competing reconnecting modes at different orientations.

In order to fit one oblique tearing mode of wavelength $\lambda$ at orientation $\theta$ inside the simulation domain, it requires $L_y\geq\lambda/\mbox{sin}\theta$ (see Appendix-2), and this wavelength needs to satisfy $2\pi/\lambda\lesssim k_c=[(1/2+b_g\mbox{tan}\theta)^2+1]^{1/2}/L$ for the unstable condition of tearing modes (i.e., $\Delta'\gtrsim 0$ calculated in the Appendix-1).
For an oblique tearing mode to grow at the optimal orientation $\theta=-13^\circ$ in the initial current sheet of $L=0.8 d_i$, it requires $L_y\gtrsim 15.5 d_i$. The fastest growing mode typically has a wavenumber around $k_c/2$, and this will require $L_y\gtrsim 31 d_i$.  Thus we choose $L_y=32 d_i$, that should provide sufficient room for the oblique tearing mode to grow at this optimal orientation if its growth is desired. This $y$-extent is eight times shorter than the primary 3D case, as illustrated by the green dashed box marked in Fig.~\ref{orientation_3D}(c). In addition, we apply a perturbation that is uniform in the $y$-direction to initiate the x-line. 

\begin{figure}
\includegraphics[width=9cm]{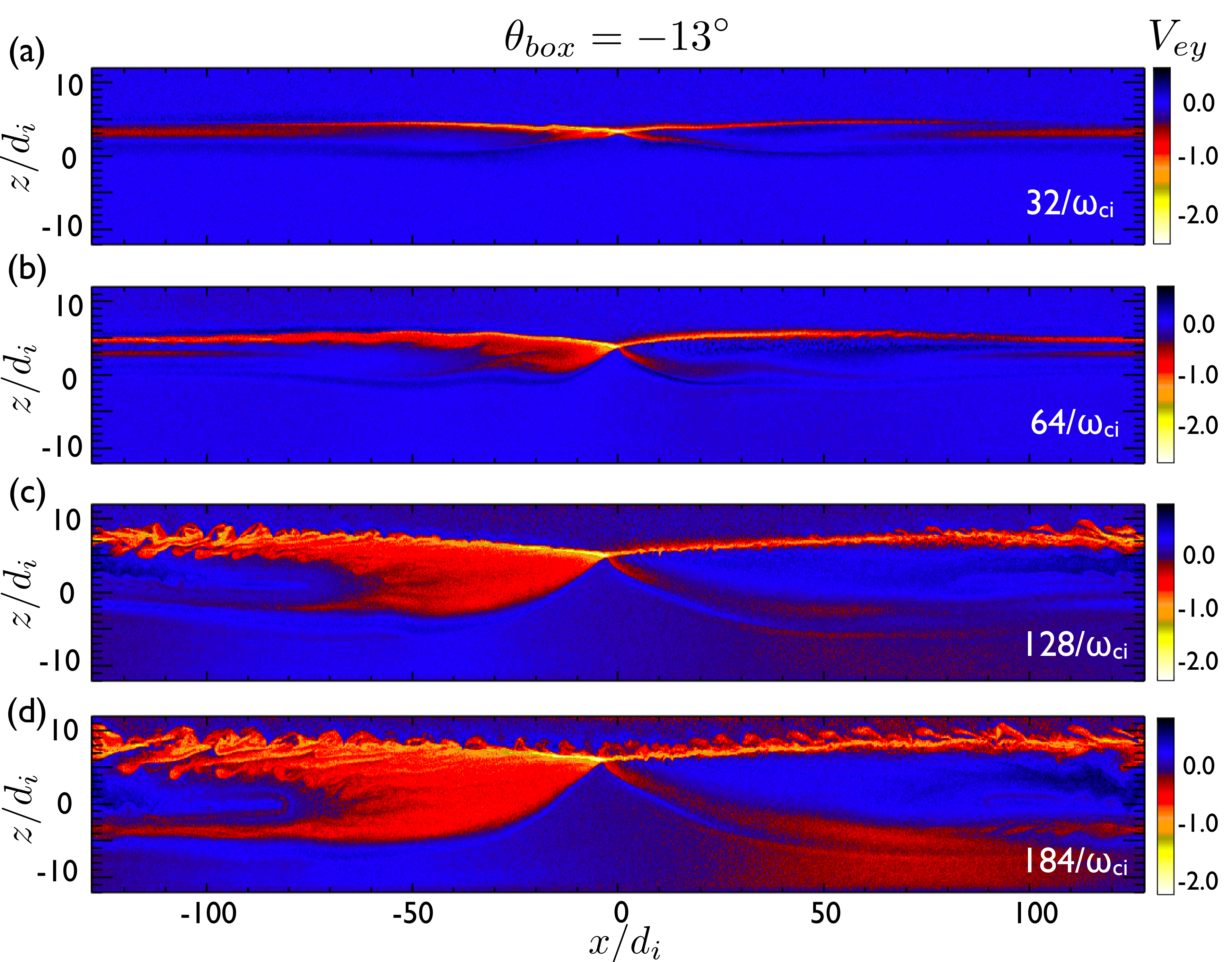} 
\caption {The evolution of reconnection in a companion 3D simulation using $L_y=32d_i$ and $\theta_{box}=-13^\circ$. The color shows the electron flow speed in the y-direction ($V_{ey}$). 
}
\label{turbulence_evolve2}
\end{figure}

\begin{figure}
\includegraphics[width=8cm]{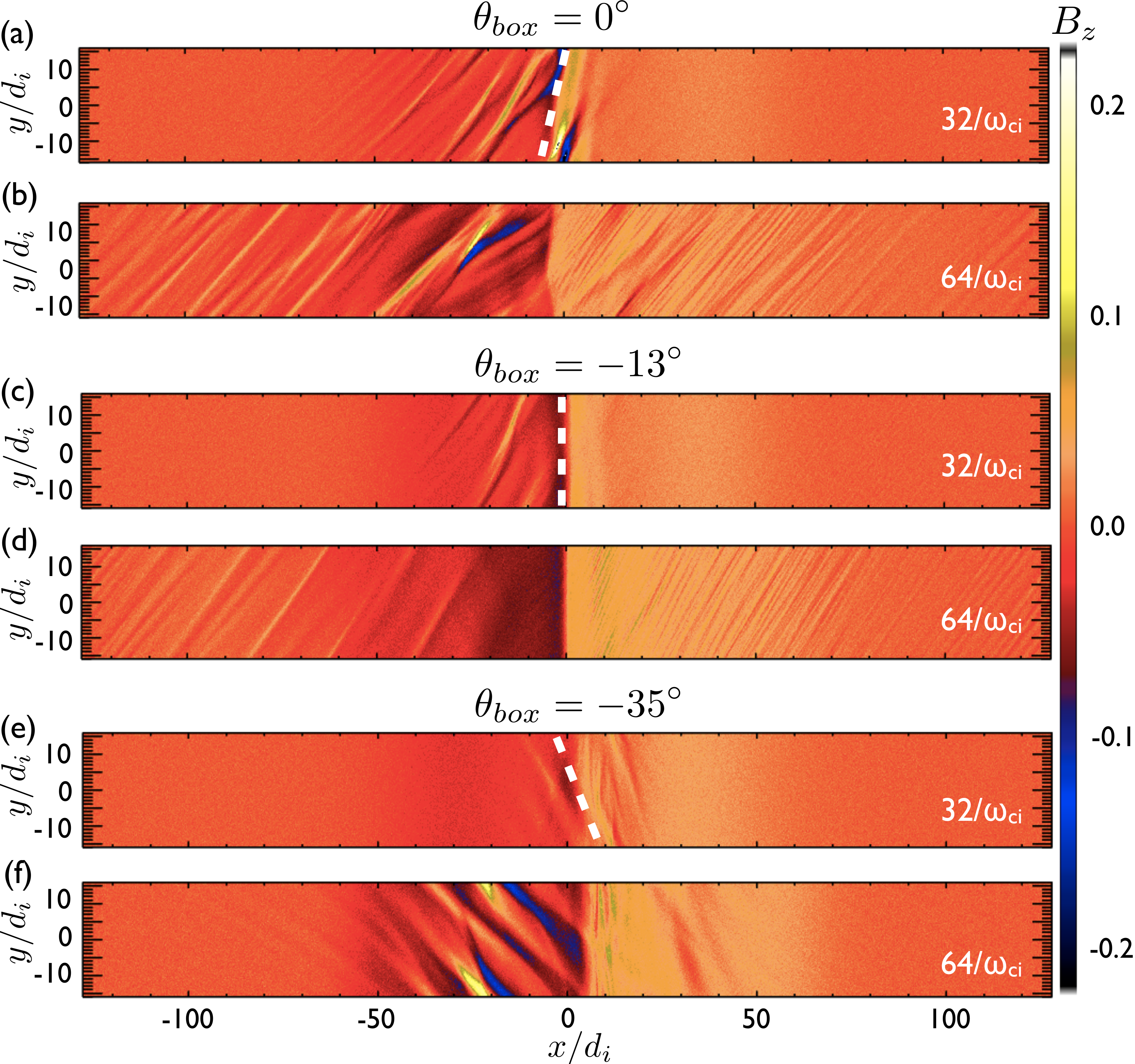} 
\caption {The $B_z$ structure in the x-y plane that contain the x-line in companion 3D simulation using $L_y=32d_i$. In (a)-(b) $\theta_{box}=0^\circ$, in (c)-(d) $\theta_{box}=-13^\circ$, in (e)-(f) $\theta_{box}=-35^\circ$. The white dashed lines mark the orientation ($\theta=-13^\circ$) favored by the local physics.
}
\label{tearing_onset}
\end{figure}

In the first case, we keep $\theta_{box}=0$. The evolution of reconnection is shown in Fig.~\ref{turbulence_evolve}. The color shows the electron velocity $V_{ey}$. The most pronounced feature is the turbulence in (c) and (d), that is absent in the large 3D case (Fig.~\ref{orientation_3D}). Here we explain what gives rise to this turbulence. First of all, note that the primary x-line points more or less in the $y$-direction as initiated by the perturbation and soon enforced by the periodic boundary condition in the y-direction. However, secondary tearing modes emerge on top of the primary x-line in Fig.~\ref{turbulence_evolve} (a) and (b). These tearing modes are oblique to the primary x-line in the y-direction, as illustrated by the $B_z$ structure on the $x-y$ plane in Fig.~\ref{tearing_onset} (a) and (b). Not too surprisingly, this structure is parallel to the optimal orientation at $\theta=-13^\circ$ as marked by the white dashed line. The system radiates secondary tearing modes to adjust the orientation, but this attempt is destined to fail because of the large-scale orientation enforced by the periodic $y$-boundary. 
The fast-streaming electrons resonated by tearing modes form intense electric current, which needs to close itself since $\nabla \cdot {\bf J}\simeq \nabla\cdot ({\nabla\times c^2{\bf B}/4\pi})\simeq 0$ in the non-relativistic limit. The intense current structure leaves one $y$-boundary at an oblique angle will come back from the other side farther downstream, forming a tearing chain along the entire separatrix and constantly feeding complexity back to the periodic system. In contrast, the x-line and separatrix are quiet in the primary 3D case (Fig.~\ref{orientation_3D}). 

\begin{figure}
\includegraphics[width=9cm]{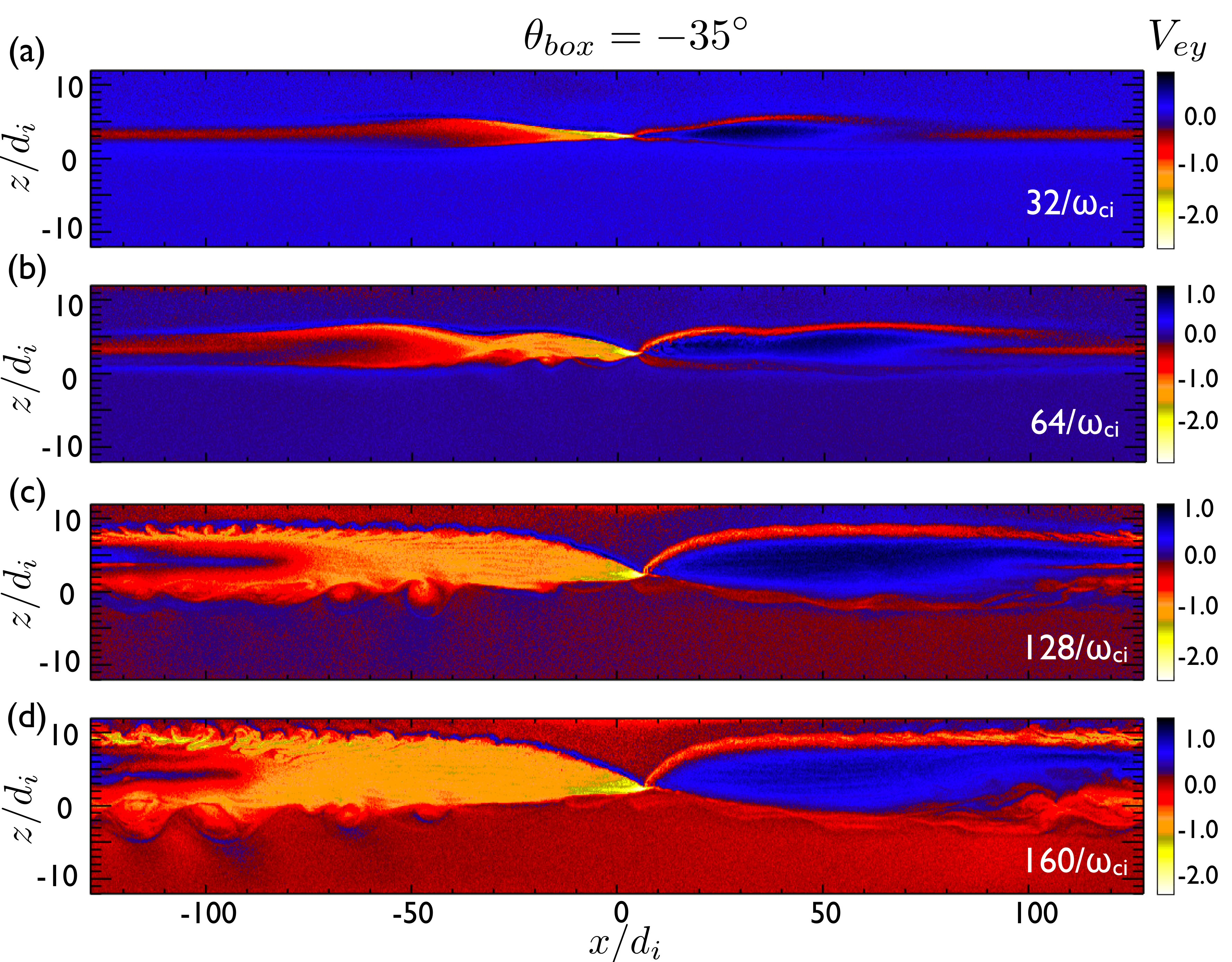} 
\caption {The evolution of reconnection in a companion 3D simulation using $L_y=32d_i$ and $\theta_{box}=-35^\circ$. The color shows the electron flow speed in the y-direction ($V_{ey}$). 
}
\label{turbulence_evolve3}
\end{figure}

In the second case, we rotate the simulation box to $\theta_{box}=-13^\circ$ so that the $y$-axis is along the optimal x-line orientation. The evolution is shown in Fig.~\ref{turbulence_evolve2}. A secondary tearing mode appears in Fig.~\ref{turbulence_evolve2} (a) and soon disappears in the outflow. This secondary tearing mode forms structure parallel to the $y$-direction, as expected, and it is easier to be advected out coherently and be merged in the outflow. The reconnection x-line is thus considerably less turbulent. Oblique modes of smaller spatial-scale later develop along the separatrix further downstream (Fig.~\ref{turbulence_evolve2}(c)). These modes could be lower-hybrid drift modes or weaker oblique tearing modes. They eventually spread out and reach the x-line (Fig.~\ref{turbulence_evolve2}(d)), perhaps, due to the combination of the $x$- and $y$- periodic boundaries. 
In the third case shown in Fig.~\ref{turbulence_evolve3}, we rotate the simulation box to $\theta_{box}=-35^\circ$. Secondary tearing modes emerge and linger around the x-line. This case further confirms that the secondary tearing modes do emerge along the optimal x-line orientation, as shown in Fig.~\ref{tearing_onset} (e) and (f). 
Note that since the primary outflow speed driven by the pre-selected x-line only varies as a function of the x location, segments on an oblique structure at different x locations are thus advected in different speeds (before the entire structure enters the region of a uniform Alfv\'enic outflow). Thus the tilt angle of the oblique structure can become larger further downstream.

In short, these numerical experiments suggest that when the primary x-line is forced to point at an orientation not favored by the local physics, the system radiates oblique tearing modes to adjust itself. The resulting oblique structure makes reconnection difficult to regain a coherent quasi-2D structure inside a small periodic box.

\subsection{7. Summary and Discussion} 
We studied the x-line orientation and its stability using particle-in-cell simulations, showing that the x-line in a large 3D system (i.e., a proxy of an open system) proceeds along the orientation that maximizes the reconnection rate. The resulting diffusion region is laminar and the non-gyrotropic feature of the pressure tensor breaks the frozen-in condition. In contrast, when the x-line is externally forced to misalign with this optimal orientation, secondary oblique tearing modes develop to adjust the orientation. Inside a small periodic system, the oblique structure can hardly be expelled and merged. The fast-streaming electrons resonated by tearing modes quickly spread over the entire system, constantly feeding complex structure back to the periodic system and leading to turbulence. Based on these numerical experiments, we conclude that the reconnection x-line needs not be as turbulent as observed in small periodic simulations. 


At Earth's magnetopause, a global reconnection-line that misaligns with the optimal orientation favored by the local physics is expected to radiate secondary oblique tearing modes. However, the relatively large system may provide a sufficient room for the x-line to adjust its orientation and to resume its natural, quieter, state. To accurately model this reaction would require more realistic initial conditions, boundary conditions, and global external drives that are not yet feasible in a full PIC simulation. One possibility is that a misaligned reconnection-line will break up into smaller segments, which each are ideally aligned. This could explain localized bursts of reconnection in connection with flux transfer events (FTEs). Note that the turbulence driven by the lower-hybrid drift instability (LHDI) was discussed in MMS observation \citep{ergun16a,graham17a} and the associated event studies using 3D particle-in-cell simulations \citep{price16a,le17a}. 
For the parameters studied in this case, the LHDI appears to be relatively weak at the x-line as shown in Fig.~\ref{orientation_3D}(a) or \ref{Ohms}(a). The effect of LHDI on the x-line is not the focus in this work but the potential boundary effect inside a small periodic system also deserves future investigation.
Note that this work does not imply that the generation of secondary magnetic islands is entirely excluded when the x-line develops along the optimal orientation. For instance, magnetic islands were observed in the vicinity of the x-line during tail reconnection \citep{RWang15a,RWang10a}. Instead, this work suggests that an x-line is inclined to generate secondary tearing modes when it misaligns with the optimal orientation.

We emphasize that an important nature of magnetic reconnection is revealed in this 3D simulation; the comparison between the observed orientation and companion 2D simulations in Fig.~\ref{orientation_2D}(a) shows that reconnection tends to proceed at or, at least, near the maximal reconnection rate. This fact can be crucial for the explanation of the fast rate value of order 0.1; a recent model \citep{yhliu17a, yhliu18a} suggests that the reconnection rate profile as a function of the opening angle made by the upstream magnetic field is relatively flat near this optimal state, and it has a value of order 0.1.

In summary, this study advances our understanding of the 3D orientation and stability of the asymmetric reconnection x-line. This result could help interpret the local geometry of reconnection events observed by Magnetospheric Multiscale Mission (MMS) and, perhaps, help determine an appropriate LMN coordinate. The question we are exploring is also relevant to the upcoming ESA-CAS joint mission, Solar wind Magnetosphere Ionosphere Link Explorer (SMILE), that will study the development of reconnection-lines at Earth's magnetopause using x-ray and UV imagers.

\section{Appendix}
{\it 1. Collisionless tearing growth rate-} In addition to obtaining an optimal orientation by maximizing the reconnection rate, it is also interesting to consider the competition of linear tearing modes that lead to spontaneous reconnection.

We consider the collisionless tearing stability of this configuration for an arbitrary wavevector  ${\bf k}=k_x\hat{{\bf x}}+k_y\hat{{\bf y}}$ corresponding to oblique angle $\theta \equiv \tan^{-1}(k_y/k_x)$ and resonance surface   
 $z_s=-L\times\mbox{arctanh}(1/2+b_g\mbox{tan}\theta)+3d_i$ at $F \equiv {\bf k \cdot B}=0$.  In the outer region, the magnetohydrodynamic model is used to obtain an eigenmode equation \citep{furth63a} of the form $\tilde{\psi}'' =(k^2+F''/F)\tilde{\psi}$, where $\tilde{\psi}(z)$ is the perturbed flux function at the oblique plane and $k^2 \equiv k_x^2 +k_y^2$. By combining the approximate solutions for $kL \ll 1$ and $kL \gg 1$ in the same manner as in Ref. \cite{baalrud12a}, we get the drive for tearing perturbations \citep{furth63a} $\Delta'\equiv \lim_{\epsilon\rightarrow 0} (1/\tilde{\psi})[d\tilde\psi/dz]_{z_s-\epsilon}^{z_s+\epsilon}\simeq (\alpha^2/k)(F^{-2}_{-\infty}+F^{-2}_{\infty})-2k$ where $\alpha\equiv(dF/dz)_{z=z_s}$. Plugging in our configuration, it gives
\begin{equation}
\Delta' \simeq \frac{2[(1/2+b_g\mbox{tan}\theta)^2+1]}{kL^2}-2k.
\label{dprime_eqn}
\end{equation}
The upper bound of the unstable wavenumber is $k_cL\lesssim [(1/2+b_g\mbox{tan}\theta)^2+1]^{1/2}$. Using the standard matching approach \citep{drake77a, daughton11a} to the kinetic resonance layer gives
\begin{equation}
\gamma \simeq \frac{d_e^2\Delta'}{l_s}kv_{the},
\label{gamma}
\end{equation}
where $v_{the} \equiv (2 T_e/m_e)^{1/2} $ is the electron thermal speed and $d_e \equiv c/\omega_{pe}$ is the local electron inertial length at the resonant surface. $l_s$ is the scale length of the magnetic shear defined in $k_{\parallel} = {\bf k} \cdot {\bf B}/|{\bf B}|\approx [\partial ({\bf k}\cdot{\bf B}/|{\bf B}|)/\partial z]_{z=z_s} (z-z_s) \equiv k (z-z_s)/l_s$. It is derived to be
 \begin{displaymath}
  l_s= \frac{L b_g (1+\mbox{tan}^2\theta)^{1/2}}{[1-(1/2+b_g\mbox{tan}\theta)^2]\mbox{cos}\theta}.
\end{displaymath} 
The dominant mode typically has a wavelength $kL\sim k_cL/2$ and it is roughly $0.5$. Based on this wavenumber ($k_c/2$), the growth rate at the different oblique angle is shown by the dashed cyan curve in Fig.~\ref{orientation_2D}(b), which has a maximum at $\theta \simeq -8^\circ$.

The width of the resonant surface $\Delta$ is determined by the resonant condition \citep{drake77a} $\gamma \simeq k_\| v_{the}=(k\Delta/l_s) v_{the}$, and it should be limited by the thickness of the current sheet $L$. Thus, by comparing with Eq.~(\ref{gamma}) we can derive $\Delta'd_e\simeq \Delta/d_e \leq L/d_e$. For $L/d_e\leq 1$, we have $\Delta'd_e \leq 1$. On the other hand,  Eq.~(\ref{dprime_eqn}) with $k\simeq k_c/2$ suggests $\Delta'd_e \simeq 1/(L/d_e)\geq 1$ in the same limit. It is thus clear that the theory breaks down for a narrow sheet $L/d_e \leq 1$. As a quick remedy, we argue $\Delta'd_e\simeq 1$ and thus $\gamma \simeq d_ekv_{the}/l_s$ for $L/d_e \leq 1$. This modified rate for a $d_e$-scale sheet is plotted as the orange dashed curve in Fig.~\ref{orientation_2D}(b), which has a maximum at $\theta\simeq -13.8^\circ$, comparable to the oblique angle ($\simeq -13^\circ$) of the dominant mode observed in a $d_e$-scale sheet \citep{yhliu15b}.

\begin{figure}
\includegraphics[width=8cm]{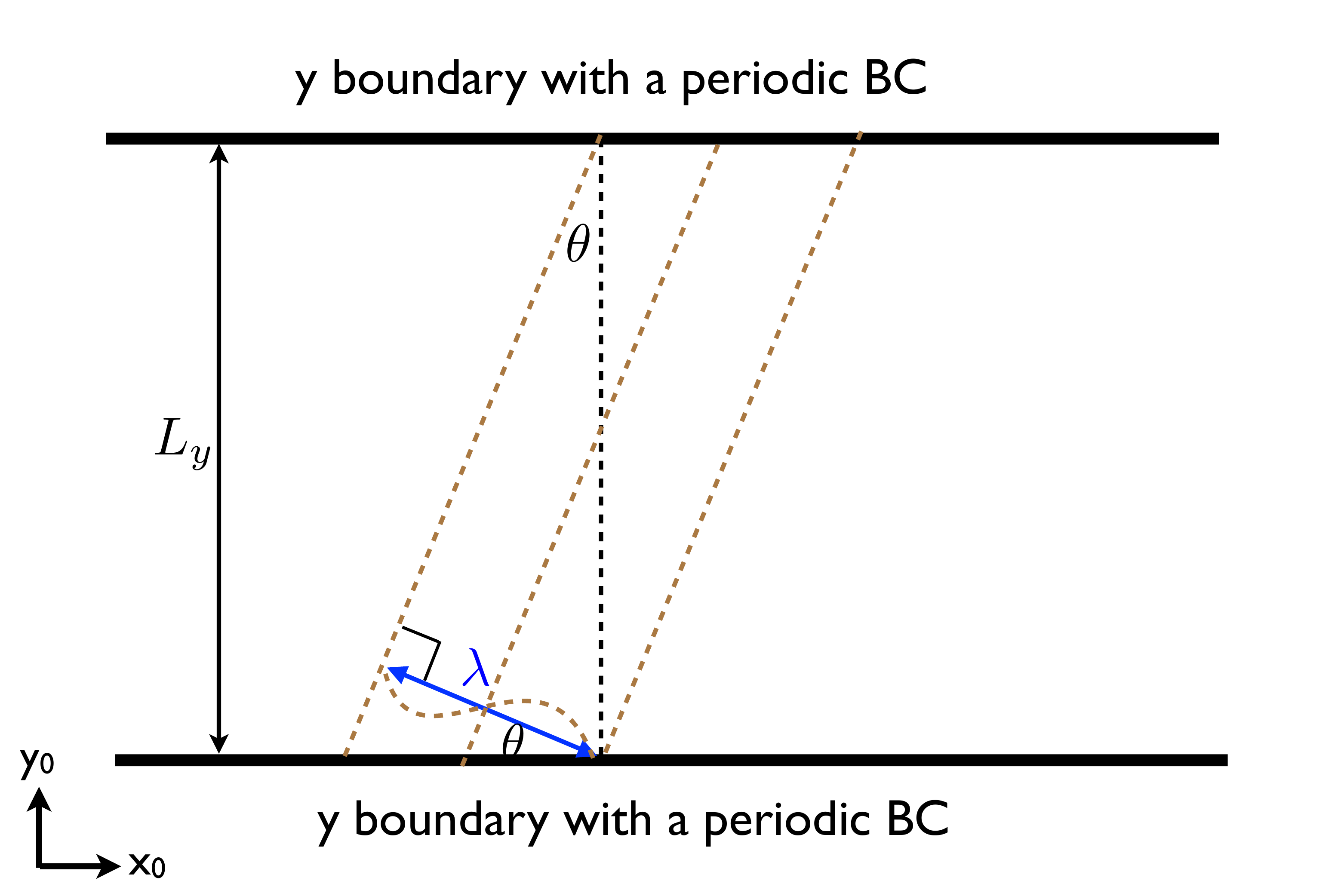} 
\caption {The system size $L_y$ that perfectly fit an oblique mode of angle $\theta$ and wavelength $\lambda$.
}
\label{box_size}
\end{figure}

{\it 2. The minimum box size required for an oblique mode-}
To perfectly fit an oblique mode of angle $\theta$ and wavelength $\lambda$ inside a box of periodic y-boundary, as shown in Fig.~\ref{box_size}, it requires $L_y=N\lambda/\mbox{sin}\theta$ where $N$ is a positive integer. For $L_y > \lambda/\mbox{sin}\theta$, the mode can at least partially manifest its orientation. For $L_y < \lambda/\mbox{sin}\theta$, such a mode is impossible to grow due to the effect of the periodic boundary.

\acknowledgments Y.-H. Liu is supported by NASA grant NNX16AG75G. M. Hesse acknowledges support by the Research Council of Norway/CoE under contract 223252/F50 and by NASA's MMS mission. Simulations is supported by NSF- Petascale Computing Resource Allocation project no.~ACI1640768, NASA Advanced Supercomputing and NERSC Advanced Supercomputing. The large date generated by peta-scale PIC simulations can hardly be made publicly available. Interested researchers are welcome to contact the leading author for subset of the data archived in computational centers. 
A movie that shows the evolution of the x-line can be found in the supporting information.


\begin{thebibliography}{53}                                                                                                                  
\expandafter\ifx\csname natexlab\endcsname\relax\def\natexlab#1{#1}\fi
\expandafter\ifx\csname bibnamefont\endcsname\relax
  \def\bibnamefont#1{#1}\fi
\expandafter\ifx\csname bibfnamefont\endcsname\relax
  \def\bibfnamefont#1{#1}\fi
\expandafter\ifx\csname citenamefont\endcsname\relax
  \def\citenamefont#1{#1}\fi
\expandafter\ifx\csname url\endcsname\relax
  \def\url#1{\texttt{#1}}\fi
\expandafter\ifx\csname urlprefix\endcsname\relax\def\urlprefix{URL }\fi
\providecommand{\bibinfo}[2]{#2}
\providecommand{\eprint}[2][]{\url{#2}}

\bibitem[{\citenamefont{Dungey}(1961)}]{dungey61a}
\bibinfo{author}{\bibfnamefont{J.}~\bibnamefont{Dungey}},
  \bibinfo{journal}{Phys. Rev. Lett.} \textbf{\bibinfo{volume}{6}},
  \bibinfo{pages}{47} (\bibinfo{year}{1961}).

\bibitem[{\citenamefont{Trattner et~al.}(2004)\citenamefont{Trattner, Fuselier,
  and Petrinec}}]{trattner04a}
\bibinfo{author}{\bibfnamefont{K.~J.} \bibnamefont{Trattner}},
  \bibinfo{author}{\bibfnamefont{S.~A.} \bibnamefont{Fuselier}},
  \bibnamefont{and} \bibinfo{author}{\bibfnamefont{S.~M.}
  \bibnamefont{Petrinec}}, \bibinfo{journal}{J. Geophys. Res}
  \textbf{\bibinfo{volume}{109}}, \bibinfo{pages}{A03219}
  (\bibinfo{year}{2004}).

\bibitem[{\citenamefont{Dunlop et~al.}(2011{\natexlab{a}})\citenamefont{Dunlop,
  Zhang, Bogdanova, Trattner, Pu, Hasegawa, Berchem, Taylor, Volwerk, Eastwood
  et~al.}}]{dunlop11a}
\bibinfo{author}{\bibfnamefont{M.~W.} \bibnamefont{Dunlop}},
  \bibinfo{author}{\bibfnamefont{Q.~H.} \bibnamefont{Zhang}},
  \bibinfo{author}{\bibfnamefont{Y.~V.} \bibnamefont{Bogdanova}},
  \bibinfo{author}{\bibfnamefont{K.~J.} \bibnamefont{Trattner}},
  \bibinfo{author}{\bibfnamefont{Z.}~\bibnamefont{Pu}},
  \bibinfo{author}{\bibfnamefont{H.}~\bibnamefont{Hasegawa}},
  \bibinfo{author}{\bibfnamefont{J.}~\bibnamefont{Berchem}},
  \bibinfo{author}{\bibfnamefont{M.~G. G.~T.} \bibnamefont{Taylor}},
  \bibinfo{author}{\bibfnamefont{M.}~\bibnamefont{Volwerk}},
  \bibinfo{author}{\bibfnamefont{J.~P.} \bibnamefont{Eastwood}},
  \bibnamefont{et~al.}, \bibinfo{journal}{Ann. Geophys.}
  \textbf{\bibinfo{volume}{29}}, \bibinfo{pages}{1683}
  (\bibinfo{year}{2011}{\natexlab{a}}).

\bibitem[{\citenamefont{Phan et~al.}(2006)\citenamefont{Phan, Gosling, Davis,
  Skoug, {\O}ieroset, Lin, Lepping, McComas, Smith, R\`eme et~al.}}]{phan06a}
\bibinfo{author}{\bibfnamefont{T.~D.} \bibnamefont{Phan}},
  \bibinfo{author}{\bibfnamefont{J.~T.} \bibnamefont{Gosling}},
  \bibinfo{author}{\bibfnamefont{M.~S.} \bibnamefont{Davis}},
  \bibinfo{author}{\bibfnamefont{R.~M.} \bibnamefont{Skoug}},
  \bibinfo{author}{\bibfnamefont{M.}~\bibnamefont{{\O}ieroset}},
  \bibinfo{author}{\bibfnamefont{R.~P.} \bibnamefont{Lin}},
  \bibinfo{author}{\bibfnamefont{R.~P.} \bibnamefont{Lepping}},
  \bibinfo{author}{\bibfnamefont{D.~J.} \bibnamefont{McComas}},
  \bibinfo{author}{\bibfnamefont{C.~W.} \bibnamefont{Smith}},
  \bibinfo{author}{\bibfnamefont{H.}~\bibnamefont{R\`eme}},
  \bibnamefont{et~al.}, \bibinfo{journal}{Nature}
  \textbf{\bibinfo{volume}{439}}, \bibinfo{pages}{175} (\bibinfo{year}{2006}).

\bibitem[{\citenamefont{Pu et~al.}(2007)\citenamefont{Pu, Zhang, Wang, Wang,
  Zhou, Dunlop, Xie, Xiao, Zong, Fu et~al.}}]{pu07a}
\bibinfo{author}{\bibfnamefont{Z.~Y.} \bibnamefont{Pu}},
  \bibinfo{author}{\bibfnamefont{X.~G.} \bibnamefont{Zhang}},
  \bibinfo{author}{\bibfnamefont{X.~G.} \bibnamefont{Wang}},
  \bibinfo{author}{\bibfnamefont{J.}~\bibnamefont{Wang}},
  \bibinfo{author}{\bibfnamefont{X.~Z.} \bibnamefont{Zhou}},
  \bibinfo{author}{\bibfnamefont{M.~W.} \bibnamefont{Dunlop}},
  \bibinfo{author}{\bibfnamefont{L.}~\bibnamefont{Xie}},
  \bibinfo{author}{\bibfnamefont{C.~J.} \bibnamefont{Xiao}},
  \bibinfo{author}{\bibfnamefont{Q.~G.} \bibnamefont{Zong}},
  \bibinfo{author}{\bibfnamefont{S.~Y.} \bibnamefont{Fu}},
  \bibnamefont{et~al.}, \bibinfo{journal}{Geophys. Res. Lett.}
  \textbf{\bibinfo{volume}{34}}, \bibinfo{pages}{L20101}
  (\bibinfo{year}{2007}).

\bibitem[{\citenamefont{Scurry et~al.}(1994)\citenamefont{Scurry, Russell, and
  Gosling}}]{scurry94a}
\bibinfo{author}{\bibfnamefont{L.}~\bibnamefont{Scurry}},
  \bibinfo{author}{\bibfnamefont{C.~T.} \bibnamefont{Russell}},
  \bibnamefont{and} \bibinfo{author}{\bibfnamefont{J.~T.}
  \bibnamefont{Gosling}}, \bibinfo{journal}{J. Geophys. Res}
  \textbf{\bibinfo{volume}{99}}, \bibinfo{pages}{14815} (\bibinfo{year}{1994}).

\bibitem[{\citenamefont{Trattner et~al.}(2007)\citenamefont{Trattner, Mulcock,
  Petrinec, and Fuselier}}]{trattner07a}
\bibinfo{author}{\bibfnamefont{K.~J.} \bibnamefont{Trattner}},
  \bibinfo{author}{\bibfnamefont{J.~S.} \bibnamefont{Mulcock}},
  \bibinfo{author}{\bibfnamefont{S.~M.} \bibnamefont{Petrinec}},
  \bibnamefont{and} \bibinfo{author}{\bibfnamefont{S.~A.}
  \bibnamefont{Fuselier}}, \bibinfo{journal}{J. Geophys. Res}
  \textbf{\bibinfo{volume}{112}}, \bibinfo{pages}{A01201}
  (\bibinfo{year}{2007}).

\bibitem[{\citenamefont{Fear et~al.}(2012)\citenamefont{Fear, Palmroth, and
  Milan}}]{fear12a}
\bibinfo{author}{\bibfnamefont{R.~C.} \bibnamefont{Fear}},
  \bibinfo{author}{\bibfnamefont{M.}~\bibnamefont{Palmroth}}, \bibnamefont{and}
  \bibinfo{author}{\bibfnamefont{S.~E.} \bibnamefont{Milan}},
  \bibinfo{journal}{J. Geophys. Res} \textbf{\bibinfo{volume}{117}},
  \bibinfo{pages}{A04202} (\bibinfo{year}{2012}).

\bibitem[{\citenamefont{Dunlop et~al.}(2011{\natexlab{b}})\citenamefont{Dunlop,
  Zhang, Bogdanova, Lockwood, Pu, Hasegawa, Wang, Taylor, Berchem, Lavraud
  et~al.}}]{dunlop11b}
\bibinfo{author}{\bibfnamefont{M.~W.} \bibnamefont{Dunlop}},
  \bibinfo{author}{\bibfnamefont{Q.~H.} \bibnamefont{Zhang}},
  \bibinfo{author}{\bibfnamefont{Y.~V.} \bibnamefont{Bogdanova}},
  \bibinfo{author}{\bibfnamefont{M.}~\bibnamefont{Lockwood}},
  \bibinfo{author}{\bibfnamefont{Z.}~\bibnamefont{Pu}},
  \bibinfo{author}{\bibfnamefont{H.}~\bibnamefont{Hasegawa}},
  \bibinfo{author}{\bibfnamefont{J.}~\bibnamefont{Wang}},
  \bibinfo{author}{\bibfnamefont{M.~G. G.~T.} \bibnamefont{Taylor}},
  \bibinfo{author}{\bibfnamefont{J.}~\bibnamefont{Berchem}},
  \bibinfo{author}{\bibfnamefont{B.}~\bibnamefont{Lavraud}},
  \bibnamefont{et~al.}, \bibinfo{journal}{Phys. Rev. Lett.}
  \textbf{\bibinfo{volume}{107}}, \bibinfo{pages}{025004}
  (\bibinfo{year}{2011}{\natexlab{b}}).

\bibitem[{\citenamefont{Wild et~al.}(2007)\citenamefont{Wild, Milan, Davies,
  Dunlop, Wright, Carr, Balogh, R{\'e}me, Fazakerley, and
  Marchaudon}}]{wild07a}
\bibinfo{author}{\bibfnamefont{J.~A.} \bibnamefont{Wild}},
  \bibinfo{author}{\bibfnamefont{S.~E.} \bibnamefont{Milan}},
  \bibinfo{author}{\bibfnamefont{J.~A.} \bibnamefont{Davies}},
  \bibinfo{author}{\bibfnamefont{M.~W.} \bibnamefont{Dunlop}},
  \bibinfo{author}{\bibfnamefont{D.~M.} \bibnamefont{Wright}},
  \bibinfo{author}{\bibfnamefont{C.~M.} \bibnamefont{Carr}},
  \bibinfo{author}{\bibfnamefont{A.}~\bibnamefont{Balogh}},
  \bibinfo{author}{\bibfnamefont{H.}~\bibnamefont{R{\'e}me}},
  \bibinfo{author}{\bibfnamefont{A.~N.} \bibnamefont{Fazakerley}},
  \bibnamefont{and}
  \bibinfo{author}{\bibfnamefont{A.}~\bibnamefont{Marchaudon}},
  \bibinfo{journal}{Ann. Geophys.} \textbf{\bibinfo{volume}{25}},
  \bibinfo{pages}{219} (\bibinfo{year}{2007}).

\bibitem[{\citenamefont{Kawano and Russell}(2005)}]{kawano05a}
\bibinfo{author}{\bibfnamefont{H.}~\bibnamefont{Kawano}} \bibnamefont{and}
  \bibinfo{author}{\bibfnamefont{C.~T.} \bibnamefont{Russell}},
  \bibinfo{journal}{J. Geophys. Res.} \textbf{\bibinfo{volume}{110}},
  \bibinfo{pages}{A07217} (\bibinfo{year}{2005}).

\bibitem[{\citenamefont{Daly et~al.}(1984)\citenamefont{Daly, Saunders,
  Rijnbeek, Sckopke, and Russell}}]{daly84a}
\bibinfo{author}{\bibfnamefont{P.~W.} \bibnamefont{Daly}},
  \bibinfo{author}{\bibfnamefont{M.~A.} \bibnamefont{Saunders}},
  \bibinfo{author}{\bibfnamefont{R.~P.} \bibnamefont{Rijnbeek}},
  \bibinfo{author}{\bibfnamefont{N.}~\bibnamefont{Sckopke}}, \bibnamefont{and}
  \bibinfo{author}{\bibfnamefont{C.~T.} \bibnamefont{Russell}},
  \bibinfo{journal}{J. Geophys. Res} \textbf{\bibinfo{volume}{89}},
  \bibinfo{pages}{3843} (\bibinfo{year}{1984}).

\bibitem[{\citenamefont{Komar et~al.}(2013)\citenamefont{Komar, Cassak,
  Dorelli, Glocer, and Kuznetsova}}]{komar13a}
\bibinfo{author}{\bibfnamefont{C.~M.} \bibnamefont{Komar}},
  \bibinfo{author}{\bibfnamefont{P.~A.} \bibnamefont{Cassak}},
  \bibinfo{author}{\bibfnamefont{J.~C.} \bibnamefont{Dorelli}},
  \bibinfo{author}{\bibfnamefont{A.}~\bibnamefont{Glocer}}, \bibnamefont{and}
  \bibinfo{author}{\bibfnamefont{M.~M.} \bibnamefont{Kuznetsova}},
  \bibinfo{journal}{J. Geophys. Res} \textbf{\bibinfo{volume}{118}},
  \bibinfo{pages}{4998} (\bibinfo{year}{2013}).

\bibitem[{\citenamefont{Alexeev et~al.}(1998)\citenamefont{Alexeev, Sibeck, and
  Bobrovnikov}}]{alexeev98a}
\bibinfo{author}{\bibfnamefont{I.~I.} \bibnamefont{Alexeev}},
  \bibinfo{author}{\bibfnamefont{D.~G.} \bibnamefont{Sibeck}},
  \bibnamefont{and} \bibinfo{author}{\bibfnamefont{S.~Y.}
  \bibnamefont{Bobrovnikov}}, \bibinfo{journal}{J. Geophys. Res}
  \textbf{\bibinfo{volume}{103}}, \bibinfo{pages}{6675} (\bibinfo{year}{1998}).

\bibitem[{\citenamefont{Papadopoulos et~al.}(1999)\citenamefont{Papadopoulos,
  Goodrich, Wiltberger, Lopez, and Lyon}}]{papadopoulos99a}
\bibinfo{author}{\bibfnamefont{K.}~\bibnamefont{Papadopoulos}},
  \bibinfo{author}{\bibfnamefont{C.}~\bibnamefont{Goodrich}},
  \bibinfo{author}{\bibfnamefont{M.}~\bibnamefont{Wiltberger}},
  \bibinfo{author}{\bibfnamefont{R.}~\bibnamefont{Lopez}}, \bibnamefont{and}
  \bibinfo{author}{\bibfnamefont{J.}~\bibnamefont{Lyon}},
  \bibinfo{journal}{Phys. Chem. Earth Part C} \textbf{\bibinfo{volume}{24}},
  \bibinfo{pages}{189} (\bibinfo{year}{1999}).

\bibitem[{\citenamefont{Cowley}(1973)}]{cowley73a}
\bibinfo{author}{\bibfnamefont{S.~W.~H.} \bibnamefont{Cowley}},
  \bibinfo{journal}{Radio Sci.} \textbf{\bibinfo{volume}{8}},
  \bibinfo{pages}{903} (\bibinfo{year}{1973}).

\bibitem[{\citenamefont{Siscoe et~al.}(2001)\citenamefont{Siscoe, Erickson,
  Sonnerup, Maynard, Siebert, Weimer, and White}}]{siscoe01a}
\bibinfo{author}{\bibfnamefont{G.~L.} \bibnamefont{Siscoe}},
  \bibinfo{author}{\bibfnamefont{G.~M.} \bibnamefont{Erickson}},
  \bibinfo{author}{\bibfnamefont{B.~U.~{\"O}.} \bibnamefont{Sonnerup}},
  \bibinfo{author}{\bibfnamefont{N.~C.} \bibnamefont{Maynard}},
  \bibinfo{author}{\bibfnamefont{K.~D.} \bibnamefont{Siebert}},
  \bibinfo{author}{\bibfnamefont{D.~R.} \bibnamefont{Weimer}},
  \bibnamefont{and} \bibinfo{author}{\bibfnamefont{W.~W.} \bibnamefont{White}},
  \bibinfo{journal}{J. Geophys. Res} \textbf{\bibinfo{volume}{106}},
  \bibinfo{pages}{13015} (\bibinfo{year}{2001}).

\bibitem[{\citenamefont{Dorelli et~al.}(2007)\citenamefont{Dorelli,
  Bhattacharjee, and Raeder}}]{dorelli07a}
\bibinfo{author}{\bibfnamefont{J.~C.} \bibnamefont{Dorelli}},
  \bibinfo{author}{\bibfnamefont{A.}~\bibnamefont{Bhattacharjee}},
  \bibnamefont{and} \bibinfo{author}{\bibfnamefont{J.}~\bibnamefont{Raeder}},
  \bibinfo{journal}{J. Geophys. Res} \textbf{\bibinfo{volume}{112}},
  \bibinfo{pages}{A02202} (\bibinfo{year}{2007}).

\bibitem[{\citenamefont{Komar et~al.}(2015)\citenamefont{Komar, Fermo, and
  Cassak}}]{komar15a}
\bibinfo{author}{\bibfnamefont{C.~M.} \bibnamefont{Komar}},
  \bibinfo{author}{\bibfnamefont{R.~L.} \bibnamefont{Fermo}}, \bibnamefont{and}
  \bibinfo{author}{\bibfnamefont{P.~A.} \bibnamefont{Cassak}},
  \bibinfo{journal}{J. Geophys. Res} \textbf{\bibinfo{volume}{120}},
  \bibinfo{pages}{276} (\bibinfo{year}{2015}).

\bibitem[{\citenamefont{Sonnerup}(1974)}]{sonnerup74a}
\bibinfo{author}{\bibfnamefont{B.~U.~{\"O}.} \bibnamefont{Sonnerup}},
  \bibinfo{journal}{J. Geophys. Res.} \textbf{\bibinfo{volume}{79}},
  \bibinfo{pages}{1546} (\bibinfo{year}{1974}).

\bibitem[{\citenamefont{Gonzalez and Mozer}(1974)}]{gonzalez74a}
\bibinfo{author}{\bibfnamefont{W.~D.} \bibnamefont{Gonzalez}} \bibnamefont{and}
  \bibinfo{author}{\bibfnamefont{F.~S.} \bibnamefont{Mozer}},
  \bibinfo{journal}{J. Geophys. Res} \textbf{\bibinfo{volume}{79}},
  \bibinfo{pages}{4186} (\bibinfo{year}{1974}).

\bibitem[{\citenamefont{Swisdak and Drake}(2007)}]{swisdak07a}
\bibinfo{author}{\bibfnamefont{M.}~\bibnamefont{Swisdak}} \bibnamefont{and}
  \bibinfo{author}{\bibfnamefont{J.~F.} \bibnamefont{Drake}},
  \bibinfo{journal}{Geophys. Res. Lett.} \textbf{\bibinfo{volume}{34}},
  \bibinfo{pages}{L11106} (\bibinfo{year}{2007}).

\bibitem[{\citenamefont{Schreier et~al.}(2010)\citenamefont{Schreier, Swisdak,
  Drake, and Cassak}}]{schreier10a}
\bibinfo{author}{\bibfnamefont{R.}~\bibnamefont{Schreier}},
  \bibinfo{author}{\bibfnamefont{M.}~\bibnamefont{Swisdak}},
  \bibinfo{author}{\bibfnamefont{J.~F.} \bibnamefont{Drake}}, \bibnamefont{and}
  \bibinfo{author}{\bibfnamefont{P.~A.} \bibnamefont{Cassak}},
  \bibinfo{journal}{Phys. Plasmas} \textbf{\bibinfo{volume}{17}},
  \bibinfo{pages}{110704} (\bibinfo{year}{2010}).

\bibitem[{\citenamefont{Hesse et~al.}(2013)\citenamefont{Hesse, Aunai,
  Zenitani, Kuznetsova, and Birn}}]{hesse13a}
\bibinfo{author}{\bibfnamefont{M.}~\bibnamefont{Hesse}},
  \bibinfo{author}{\bibfnamefont{N.}~\bibnamefont{Aunai}},
  \bibinfo{author}{\bibfnamefont{S.}~\bibnamefont{Zenitani}},
  \bibinfo{author}{\bibfnamefont{M.}~\bibnamefont{Kuznetsova}},
  \bibnamefont{and} \bibinfo{author}{\bibfnamefont{J.}~\bibnamefont{Birn}},
  \bibinfo{journal}{Phys. Plasmas} \textbf{\bibinfo{volume}{20}},
  \bibinfo{pages}{061210} (\bibinfo{year}{2013}).

\bibitem[{\citenamefont{{Yi-Hsin Liu} et~al.}(2015)\citenamefont{{Yi-Hsin Liu},
  Hesse, and Kuznetsova}}]{yhliu15b}
\bibinfo{author}{\bibnamefont{{Yi-Hsin Liu}}},
  \bibinfo{author}{\bibfnamefont{M.}~\bibnamefont{Hesse}}, \bibnamefont{and}
  \bibinfo{author}{\bibfnamefont{M.}~\bibnamefont{Kuznetsova}},
  \bibinfo{journal}{J. Geophys. Res.} \textbf{\bibinfo{volume}{120}},
  \bibinfo{pages}{7331} (\bibinfo{year}{2015}).

\bibitem[{\citenamefont{Aunai et~al.}(2016)\citenamefont{Aunai, Hesse, Lavraud,
  Dargent, and Smets}}]{aunai16a}
\bibinfo{author}{\bibfnamefont{N.}~\bibnamefont{Aunai}},
  \bibinfo{author}{\bibfnamefont{M.}~\bibnamefont{Hesse}},
  \bibinfo{author}{\bibfnamefont{B.}~\bibnamefont{Lavraud}},
  \bibinfo{author}{\bibfnamefont{J.}~\bibnamefont{Dargent}}, \bibnamefont{and}
  \bibinfo{author}{\bibfnamefont{R.}~\bibnamefont{Smets}}, \bibinfo{journal}{J.
  Plasmas Phys.} \textbf{\bibinfo{volume}{82}}, \bibinfo{pages}{535820401}
  (\bibinfo{year}{2016}).

\bibitem[{\citenamefont{Bowers et~al.}(2009)\citenamefont{Bowers, Albright,
  Yin, Daughton, Roytershteyn, Bergen, and Kwan}}]{bowers09a}
\bibinfo{author}{\bibfnamefont{K.}~\bibnamefont{Bowers}},
  \bibinfo{author}{\bibfnamefont{B.}~\bibnamefont{Albright}},
  \bibinfo{author}{\bibfnamefont{L.}~\bibnamefont{Yin}},
  \bibinfo{author}{\bibfnamefont{W.}~\bibnamefont{Daughton}},
  \bibinfo{author}{\bibfnamefont{V.}~\bibnamefont{Roytershteyn}},
  \bibinfo{author}{\bibfnamefont{B.}~\bibnamefont{Bergen}}, \bibnamefont{and}
  \bibinfo{author}{\bibfnamefont{T.}~\bibnamefont{Kwan}},
  \bibinfo{journal}{Journal of Physics: Conference Series}
  \textbf{\bibinfo{volume}{180}}, \bibinfo{pages}{012055}
  (\bibinfo{year}{2009}).

\bibitem[{\citenamefont{Aunai et~al.}(2013{\natexlab{a}})\citenamefont{Aunai,
  Hesse, Zenitani, Kuznetsova, Black, Evans, and Smets}}]{aunai13b}
\bibinfo{author}{\bibfnamefont{N.}~\bibnamefont{Aunai}},
  \bibinfo{author}{\bibfnamefont{M.}~\bibnamefont{Hesse}},
  \bibinfo{author}{\bibfnamefont{S.}~\bibnamefont{Zenitani}},
  \bibinfo{author}{\bibfnamefont{M.}~\bibnamefont{Kuznetsova}},
  \bibinfo{author}{\bibfnamefont{C.}~\bibnamefont{Black}},
  \bibinfo{author}{\bibfnamefont{R.}~\bibnamefont{Evans}}, \bibnamefont{and}
  \bibinfo{author}{\bibfnamefont{R.}~\bibnamefont{Smets}},
  \bibinfo{journal}{Phys. Plasmas} \textbf{\bibinfo{volume}{20}},
  \bibinfo{pages}{022902} (\bibinfo{year}{2013}{\natexlab{a}}).

\bibitem[{\citenamefont{Pritchett}(2008)}]{pritchett08a}
\bibinfo{author}{\bibfnamefont{P.~L.} \bibnamefont{Pritchett}},
  \bibinfo{journal}{J. Geophys. Res.} \textbf{\bibinfo{volume}{113}},
  \bibinfo{pages}{A06210} (\bibinfo{year}{2008}).

\bibitem[{\citenamefont{{Yi-Hsin Liu} et~al.}(2018)\citenamefont{{Yi-Hsin Liu},
  Hesse, Cassak, Shay, Wang, and {L.-J. Chen}}}]{yhliu18a}
\bibinfo{author}{\bibnamefont{{Yi-Hsin Liu}}},
  \bibinfo{author}{\bibfnamefont{M.}~\bibnamefont{Hesse}},
  \bibinfo{author}{\bibfnamefont{P.~A.} \bibnamefont{Cassak}},
  \bibinfo{author}{\bibfnamefont{M.~A.} \bibnamefont{Shay}},
  \bibinfo{author}{\bibfnamefont{S.}~\bibnamefont{Wang}}, \bibnamefont{and}
  \bibinfo{author}{\bibnamefont{{L.-J. Chen}}}, \bibinfo{journal}{Geophys. Res.
  Lett.} \textbf{\bibinfo{volume}{45}} (\bibinfo{year}{2018}).

\bibitem[{\citenamefont{Karimabadi et~al.}(2003)\citenamefont{Karimabadi,
  Daughton, Pritchett, and Krauss-Varban}}]{karimabadi03a}
\bibinfo{author}{\bibfnamefont{H.}~\bibnamefont{Karimabadi}},
  \bibinfo{author}{\bibfnamefont{W.}~\bibnamefont{Daughton}},
  \bibinfo{author}{\bibfnamefont{P.~L.} \bibnamefont{Pritchett}},
  \bibnamefont{and}
  \bibinfo{author}{\bibfnamefont{D.}~\bibnamefont{Krauss-Varban}},
  \bibinfo{journal}{J. Geophys. Res.} \textbf{\bibinfo{volume}{108}},
  \bibinfo{pages}{1400} (\bibinfo{year}{2003}).

\bibitem[{\citenamefont{Burch et~al.}(2016)\citenamefont{Burch, Tobert, Phan,
  Chen, and {Moore et al.,}}}]{burch16a}
\bibinfo{author}{\bibfnamefont{J.~L.} \bibnamefont{Burch}},
  \bibinfo{author}{\bibfnamefont{R.~B.} \bibnamefont{Tobert}},
  \bibinfo{author}{\bibfnamefont{T.}~\bibnamefont{Phan}},
  \bibinfo{author}{\bibfnamefont{L.~J.} \bibnamefont{Chen}}, \bibnamefont{and}
  \bibinfo{author}{\bibfnamefont{T.~E.} \bibnamefont{{Moore et al.,}}},
  \bibinfo{journal}{Science} \textbf{\bibinfo{volume}{10}},
  \bibinfo{pages}{1126} (\bibinfo{year}{2016}).

\bibitem[{\citenamefont{Aunai et~al.}(2013{\natexlab{b}})\citenamefont{Aunai,
  Hesse, and Kuznetsova}}]{aunai13d}
\bibinfo{author}{\bibfnamefont{N.}~\bibnamefont{Aunai}},
  \bibinfo{author}{\bibfnamefont{M.}~\bibnamefont{Hesse}}, \bibnamefont{and}
  \bibinfo{author}{\bibfnamefont{M.}~\bibnamefont{Kuznetsova}},
  \bibinfo{journal}{Phys. Plasmas} \textbf{\bibinfo{volume}{20}},
  \bibinfo{pages}{092903} (\bibinfo{year}{2013}{\natexlab{b}}).

\bibitem[{\citenamefont{Che et~al.}(2011)\citenamefont{Che, Drake, and
  Swisdak}}]{che11a}
\bibinfo{author}{\bibfnamefont{H.}~\bibnamefont{Che}},
  \bibinfo{author}{\bibfnamefont{J.~F.} \bibnamefont{Drake}}, \bibnamefont{and}
  \bibinfo{author}{\bibfnamefont{M.}~\bibnamefont{Swisdak}},
  \bibinfo{journal}{Nature} \textbf{\bibinfo{volume}{474}},
  \bibinfo{pages}{184} (\bibinfo{year}{2011}).

\bibitem[{\citenamefont{Price et~al.}(2016)\citenamefont{Price, Swisdak, Drake,
  Cassak, Dahlin, and Ergun}}]{price16a}
\bibinfo{author}{\bibfnamefont{L.}~\bibnamefont{Price}},
  \bibinfo{author}{\bibfnamefont{M.}~\bibnamefont{Swisdak}},
  \bibinfo{author}{\bibfnamefont{J.~F.} \bibnamefont{Drake}},
  \bibinfo{author}{\bibfnamefont{P.~A.} \bibnamefont{Cassak}},
  \bibinfo{author}{\bibfnamefont{J.~T.} \bibnamefont{Dahlin}},
  \bibnamefont{and} \bibinfo{author}{\bibfnamefont{R.~E.} \bibnamefont{Ergun}},
  \bibinfo{journal}{Geophys. Res. Lett.} \textbf{\bibinfo{volume}{43}},
  \bibinfo{pages}{6020} (\bibinfo{year}{2016}).

\bibitem[{\citenamefont{Le et~al.}(2017)\citenamefont{Le, Daughton, {L.-J.
  Chen}, and Egedal}}]{le17a}
\bibinfo{author}{\bibfnamefont{A.}~\bibnamefont{Le}},
  \bibinfo{author}{\bibfnamefont{W.}~\bibnamefont{Daughton}},
  \bibinfo{author}{\bibnamefont{{L.-J. Chen}}}, \bibnamefont{and}
  \bibinfo{author}{\bibfnamefont{J.}~\bibnamefont{Egedal}},
  \bibinfo{journal}{Geophys. Res. Lett.} \textbf{\bibinfo{volume}{44}},
  \bibinfo{pages}{2096} (\bibinfo{year}{2017}).

\bibitem[{\citenamefont{Le et~al.}(2018)\citenamefont{Le, Daughton, Ohia, Chen,
  {Yi-Hsin Liu}, Nystrom, and Bird}}]{le18a}
\bibinfo{author}{\bibfnamefont{A.}~\bibnamefont{Le}},
  \bibinfo{author}{\bibfnamefont{W.}~\bibnamefont{Daughton}},
  \bibinfo{author}{\bibfnamefont{O.}~\bibnamefont{Ohia}},
  \bibinfo{author}{\bibfnamefont{L.~J.} \bibnamefont{Chen}},
  \bibinfo{author}{\bibnamefont{{Yi-Hsin Liu}}},
  \bibinfo{author}{\bibfnamefont{W.}~\bibnamefont{Nystrom}}, \bibnamefont{and}
  \bibinfo{author}{\bibfnamefont{R.}~\bibnamefont{Bird}},
  \bibinfo{journal}{arXiv:1802.10205}  (\bibinfo{year}{2018}).

\bibitem[{\citenamefont{Hesse et~al.}(2016)\citenamefont{Hesse, Liu, Chen,
  Bessho, Kuznetsova, Birn, and Burch}}]{hesse16a}
\bibinfo{author}{\bibfnamefont{M.}~\bibnamefont{Hesse}},
  \bibinfo{author}{\bibfnamefont{Y.}~\bibnamefont{Liu}},
  \bibinfo{author}{\bibfnamefont{L.-J.} \bibnamefont{Chen}},
  \bibinfo{author}{\bibfnamefont{N.}~\bibnamefont{Bessho}},
  \bibinfo{author}{\bibfnamefont{M.}~\bibnamefont{Kuznetsova}},
  \bibinfo{author}{\bibfnamefont{J.}~\bibnamefont{Birn}}, \bibnamefont{and}
  \bibinfo{author}{\bibfnamefont{J.~L.} \bibnamefont{Burch}},
  \bibinfo{journal}{Geophys. Res. Lett.} \textbf{\bibinfo{volume}{43}},
  \bibinfo{pages}{2359} (\bibinfo{year}{2016}).

\bibitem[{\citenamefont{Lu et~al.}(2013)\citenamefont{Lu, Lu, Huang, Wu, and
  Wang}}]{QLu13a}
\bibinfo{author}{\bibfnamefont{Q.}~\bibnamefont{Lu}},
  \bibinfo{author}{\bibfnamefont{S.}~\bibnamefont{Lu}},
  \bibinfo{author}{\bibfnamefont{C.}~\bibnamefont{Huang}},
  \bibinfo{author}{\bibfnamefont{M.}~\bibnamefont{Wu}}, \bibnamefont{and}
  \bibinfo{author}{\bibfnamefont{S.}~\bibnamefont{Wang}},
  \bibinfo{journal}{Plasma Phys. Control. Fusion}
  \textbf{\bibinfo{volume}{55}}, \bibinfo{pages}{085019}
  (\bibinfo{year}{2013}).

\bibitem[{\citenamefont{Rager et~al.}(2018)\citenamefont{Rager, Dorelli,
  Gershman, and {V. Uritsky et. al.,}}}]{rager18a}
\bibinfo{author}{\bibfnamefont{A.~C.} \bibnamefont{Rager}},
  \bibinfo{author}{\bibfnamefont{J.~C.} \bibnamefont{Dorelli}},
  \bibinfo{author}{\bibfnamefont{D.~J.} \bibnamefont{Gershman}},
  \bibnamefont{and} \bibinfo{author}{\bibnamefont{{V. Uritsky et. al.,}}},
  \bibinfo{journal}{Geophys. Res. Lett.} \textbf{\bibinfo{volume}{45}},
  \bibinfo{pages}{578} (\bibinfo{year}{2018}).

\bibitem[{\citenamefont{Genestreti et~al.}(2018)\citenamefont{Genestreti,
  Varsani, Burch, and {P. A. Cassak et. al.,}}}]{genestreti18a}
\bibinfo{author}{\bibfnamefont{K.~J.} \bibnamefont{Genestreti}},
  \bibinfo{author}{\bibfnamefont{A.}~\bibnamefont{Varsani}},
  \bibinfo{author}{\bibfnamefont{J.~L.} \bibnamefont{Burch}}, \bibnamefont{and}
  \bibinfo{author}{\bibnamefont{{P. A. Cassak et. al.,}}},
  \bibinfo{journal}{Geophys. Res. Lett.} \textbf{\bibinfo{volume}{123}}
  (\bibinfo{year}{2018}).

\bibitem[{\citenamefont{{Yi-Hsin Liu} et~al.}(2013)\citenamefont{{Yi-Hsin Liu},
  Daughton, Karimabadi, Li, and Roytershteyn}}]{yhliu13a}
\bibinfo{author}{\bibnamefont{{Yi-Hsin Liu}}},
  \bibinfo{author}{\bibfnamefont{W.}~\bibnamefont{Daughton}},
  \bibinfo{author}{\bibfnamefont{H.}~\bibnamefont{Karimabadi}},
  \bibinfo{author}{\bibfnamefont{H.}~\bibnamefont{Li}}, \bibnamefont{and}
  \bibinfo{author}{\bibfnamefont{V.}~\bibnamefont{Roytershteyn}},
  \bibinfo{journal}{Phys. Rev. Lett.} \textbf{\bibinfo{volume}{110}},
  \bibinfo{pages}{265004} (\bibinfo{year}{2013}).

\bibitem[{\citenamefont{Cassak and Shay}(2007)}]{cassak07b}
\bibinfo{author}{\bibfnamefont{P.~A.} \bibnamefont{Cassak}} \bibnamefont{and}
  \bibinfo{author}{\bibfnamefont{M.~A.} \bibnamefont{Shay}},
  \bibinfo{journal}{Phys. Plasmas} \textbf{\bibinfo{volume}{14}},
  \bibinfo{pages}{102114} (\bibinfo{year}{2007}).

\bibitem[{\citenamefont{Birn et~al.}(2010)\citenamefont{Birn, Borovsky, Hesse,
  and Schindler}}]{birn10a}
\bibinfo{author}{\bibfnamefont{J.}~\bibnamefont{Birn}},
  \bibinfo{author}{\bibfnamefont{J.~E.} \bibnamefont{Borovsky}},
  \bibinfo{author}{\bibfnamefont{M.}~\bibnamefont{Hesse}}, \bibnamefont{and}
  \bibinfo{author}{\bibfnamefont{K.}~\bibnamefont{Schindler}},
  \bibinfo{journal}{Phys. Plasmas} \textbf{\bibinfo{volume}{17}},
  \bibinfo{pages}{052108} (\bibinfo{year}{2010}).

\bibitem[{\citenamefont{Ergun et~al.}(2016)\citenamefont{Ergun, Goodrich,
  Wilder, Holmes, Stawarz, Eriksson, Sturner, Malaspina, Usanova, Torbert
  et~al.}}]{ergun16a}
\bibinfo{author}{\bibfnamefont{R.~E.} \bibnamefont{Ergun}},
  \bibinfo{author}{\bibfnamefont{K.~A.} \bibnamefont{Goodrich}},
  \bibinfo{author}{\bibfnamefont{F.~D.} \bibnamefont{Wilder}},
  \bibinfo{author}{\bibfnamefont{J.~C.} \bibnamefont{Holmes}},
  \bibinfo{author}{\bibfnamefont{J.~E.} \bibnamefont{Stawarz}},
  \bibinfo{author}{\bibfnamefont{S.}~\bibnamefont{Eriksson}},
  \bibinfo{author}{\bibfnamefont{A.~P.} \bibnamefont{Sturner}},
  \bibinfo{author}{\bibfnamefont{D.~M.} \bibnamefont{Malaspina}},
  \bibinfo{author}{\bibfnamefont{M.~E.} \bibnamefont{Usanova}},
  \bibinfo{author}{\bibfnamefont{R.~B.} \bibnamefont{Torbert}},
  \bibnamefont{et~al.}, \bibinfo{journal}{Phys. Rev. Lett.}
  \textbf{\bibinfo{volume}{116}}, \bibinfo{pages}{235102}
  (\bibinfo{year}{2016}).

\bibitem[{\citenamefont{Graham et~al.}(2017)\citenamefont{Graham, Khotyaintsev,
  Norgren, Vaivads, Andr\'e, and {Toledo-Rendondo et al.,}}}]{graham17a}
\bibinfo{author}{\bibfnamefont{D.~B.} \bibnamefont{Graham}},
  \bibinfo{author}{\bibfnamefont{Y.~V.} \bibnamefont{Khotyaintsev}},
  \bibinfo{author}{\bibfnamefont{C.}~\bibnamefont{Norgren}},
  \bibinfo{author}{\bibfnamefont{A.}~\bibnamefont{Vaivads}},
  \bibinfo{author}{\bibfnamefont{M.}~\bibnamefont{Andr\'e}}, \bibnamefont{and}
  \bibinfo{author}{\bibfnamefont{S.}~\bibnamefont{{Toledo-Rendondo et al.,}}},
  \bibinfo{journal}{J. Geophys. Res.} \textbf{\bibinfo{volume}{122}},
  \bibinfo{pages}{517} (\bibinfo{year}{2017}).

\bibitem[{\citenamefont{Wang et~al.}(2015)\citenamefont{Wang, Lu, Nakamura,
  Huang, Du, Guo, Teh, Wu, Lu, and Wang}}]{RWang15a}
\bibinfo{author}{\bibfnamefont{R.}~\bibnamefont{Wang}},
  \bibinfo{author}{\bibfnamefont{Q.}~\bibnamefont{Lu}},
  \bibinfo{author}{\bibfnamefont{R.}~\bibnamefont{Nakamura}},
  \bibinfo{author}{\bibfnamefont{C.}~\bibnamefont{Huang}},
  \bibinfo{author}{\bibfnamefont{A.}~\bibnamefont{Du}},
  \bibinfo{author}{\bibfnamefont{F.}~\bibnamefont{Guo}},
  \bibinfo{author}{\bibfnamefont{W.}~\bibnamefont{Teh}},
  \bibinfo{author}{\bibfnamefont{M.}~\bibnamefont{Wu}},
  \bibinfo{author}{\bibfnamefont{S.}~\bibnamefont{Lu}}, \bibnamefont{and}
  \bibinfo{author}{\bibfnamefont{S.}~\bibnamefont{Wang}},
  \bibinfo{journal}{Nature Phys.} \textbf{\bibinfo{volume}{12}},
  \bibinfo{pages}{263} (\bibinfo{year}{2015}).

\bibitem[{\citenamefont{Wang et~al.}(2010)\citenamefont{Wang, Lu, Du, and
  Wang}}]{RWang10a}
\bibinfo{author}{\bibfnamefont{R.}~\bibnamefont{Wang}},
  \bibinfo{author}{\bibfnamefont{Q.}~\bibnamefont{Lu}},
  \bibinfo{author}{\bibfnamefont{A.}~\bibnamefont{Du}}, \bibnamefont{and}
  \bibinfo{author}{\bibfnamefont{S.}~\bibnamefont{Wang}},
  \bibinfo{journal}{Phys. Rev. Lett.} \textbf{\bibinfo{volume}{104}},
  \bibinfo{pages}{175003} (\bibinfo{year}{2010}).

\bibitem[{\citenamefont{{Yi-Hsin Liu} et~al.}(2017)\citenamefont{{Yi-Hsin Liu},
  Hesse, Guo, Daughton, Li, Cassak, and Shay}}]{yhliu17a}
\bibinfo{author}{\bibnamefont{{Yi-Hsin Liu}}},
  \bibinfo{author}{\bibfnamefont{M.}~\bibnamefont{Hesse}},
  \bibinfo{author}{\bibfnamefont{F.}~\bibnamefont{Guo}},
  \bibinfo{author}{\bibfnamefont{W.}~\bibnamefont{Daughton}},
  \bibinfo{author}{\bibfnamefont{H.}~\bibnamefont{Li}},
  \bibinfo{author}{\bibfnamefont{P.~A.} \bibnamefont{Cassak}},
  \bibnamefont{and} \bibinfo{author}{\bibfnamefont{M.~A.} \bibnamefont{Shay}},
  \bibinfo{journal}{Phys. Rev. Lett.} \textbf{\bibinfo{volume}{118}},
  \bibinfo{pages}{085101} (\bibinfo{year}{2017}).

\bibitem[{\citenamefont{Furth et~al.}(1963)\citenamefont{Furth, Killeen, and
  Rosenbluth}}]{furth63a}
\bibinfo{author}{\bibfnamefont{H.}~\bibnamefont{Furth}},
  \bibinfo{author}{\bibfnamefont{J.}~\bibnamefont{Killeen}}, \bibnamefont{and}
  \bibinfo{author}{\bibfnamefont{M.~N.} \bibnamefont{Rosenbluth}},
  \bibinfo{journal}{Physics of Fluids} \textbf{\bibinfo{volume}{6}},
  \bibinfo{pages}{459} (\bibinfo{year}{1963}).

\bibitem[{\citenamefont{Baalrud et~al.}(2012)\citenamefont{Baalrud,
  Bhattacharjee, and Huang}}]{baalrud12a}
\bibinfo{author}{\bibfnamefont{S.~D.} \bibnamefont{Baalrud}},
  \bibinfo{author}{\bibfnamefont{A.}~\bibnamefont{Bhattacharjee}},
  \bibnamefont{and} \bibinfo{author}{\bibfnamefont{Y.~M.} \bibnamefont{Huang}},
  \bibinfo{journal}{Phys. Plasmas} \textbf{\bibinfo{volume}{19}},
  \bibinfo{pages}{022101} (\bibinfo{year}{2012}).

\bibitem[{\citenamefont{Drake and Lee}(1977)}]{drake77a}
\bibinfo{author}{\bibfnamefont{J.~F.} \bibnamefont{Drake}} \bibnamefont{and}
  \bibinfo{author}{\bibfnamefont{Y.~C.} \bibnamefont{Lee}},
  \bibinfo{journal}{Phys. Fluids} \textbf{\bibinfo{volume}{20}},
  \bibinfo{pages}{1341} (\bibinfo{year}{1977}).

\bibitem[{\citenamefont{Daughton et~al.}(2011)\citenamefont{Daughton,
  Roytershteyn, Karimabadi, Yin, Albright, Bergen, and Bowers}}]{daughton11a}
\bibinfo{author}{\bibfnamefont{W.}~\bibnamefont{Daughton}},
  \bibinfo{author}{\bibfnamefont{V.}~\bibnamefont{Roytershteyn}},
  \bibinfo{author}{\bibfnamefont{H.}~\bibnamefont{Karimabadi}},
  \bibinfo{author}{\bibfnamefont{L.}~\bibnamefont{Yin}},
  \bibinfo{author}{\bibfnamefont{B.~J.} \bibnamefont{Albright}},
  \bibinfo{author}{\bibfnamefont{B.}~\bibnamefont{Bergen}}, \bibnamefont{and}
  \bibinfo{author}{\bibfnamefont{K.~J.} \bibnamefont{Bowers}},
  \bibinfo{journal}{Nature Physics} \textbf{\bibinfo{volume}{7}},
  \bibinfo{pages}{539} (\bibinfo{year}{2011}).

\end{thebibliography}

\end{document}